\documentclass[11pt]{article}
\usepackage{moriond,epsfig}

\usepackage{amsmath}

\def\Journal#1#2#3#4{{#1}\ {\bf #2}, #3 (#4)}

\def\NIMA{{\em Nucl.\ Instrum.\ Methods} A}
\def\NPB{{\em Nucl.\ Phys.}\ B}
\def\PLB{{\em Phys.\ Lett.}\  B}
\def\PRL{\em Phys.\ Rev.\ Lett.}
\def\PRD{{\em Phys.\ Rev.}\ D}

\def\be{\begin{equation}}
\def\ee{\end{equation}}
\def\bea{\begin{eqnarray}}
\def\eea{\end{eqnarray}}

\usepackage{relsize}
\def\babar{\mbox{\slshape B\kern-0.1em{\smaller A}\kern-0.1em
    B\kern-0.1em{\smaller A\kern-0.2em R}}}

\newcommand*{\bbar}{\ensuremath{{\overline{B}}}}
\newcommand*{\bbbar}{\ensuremath{{B\bbar}}}
\newcommand*{\UPS}{\ensuremath{{\Upsilon(4S)}}}
\newcommand*{\pip}{\ensuremath{\pi^+}}
\newcommand*{\pim}{\ensuremath{\pi^-}}
\newcommand*{\piz}{\ensuremath{\pi^0}}
\newcommand*{\rhop}{\ensuremath{{\rho^+}}}
\newcommand*{\rhoz}{\ensuremath{{\rho^0}}}
\newcommand*{\kp}{\ensuremath{{K^+}}}
\newcommand*{\km}{\ensuremath{{K^-}}}
\newcommand*{\kz}{\ensuremath{{K^0}}}
\newcommand*{\kzb}{\ensuremath{{\overline{K}{}^0}}}
\newcommand*{\ks}{\ensuremath{{K_S^0}}}
\newcommand*{\kstarp}{\ensuremath{{K^{*+}}}}
\newcommand*{\kstarz}{\ensuremath{{K^{*0}}}}
\newcommand*{\etap}{\ensuremath{{\eta^\prime}}}
\newcommand*{\bz}{\ensuremath{{B^0}}}
\newcommand*{\bzb}{\ensuremath{{\overline{B}{}^0}}}
\newcommand*{\bp}{\ensuremath{{B^+}}}
\newcommand*{\bm}{\ensuremath{{B^-}}}

\newcommand*{\dedx}{\ensuremath{{dE/dx}}}
\newcommand*{\dE}{\ensuremath{{\Delta E}}}
\newcommand*{\mb}{\ensuremath{{M_\text{bc}}}}
\newcommand*{\mes}{\ensuremath{{m_\text{ES}}}}
\newcommand*{\Eb}{\ensuremath{{E_B^\text{cms}}}}
\newcommand*{\pb}{\ensuremath{{p_B^\text{cms}}}}
\newcommand*{\Ebeam}{\ensuremath{{E_\text{beam}^\text{cms}}}}

\newcommand*{\Ns}{\ensuremath{{N_\text{S}}}}
\newcommand*{\Br}{\ensuremath{{\mathcal{B}}}}
\newcommand*{\Acp}{\ensuremath{{A_{CP}}}}

\newcommand*{\degree}{\ensuremath{{}^\circ}}
\newcommand*{\fb}{\ensuremath{\text{fb}^{-1}}} 

\newcommand*{\etal}{\textit{et al.}}

\begin{document}
\vspace*{4cm}
\title{\boldmath RARE HADRONIC $B$ DECAYS AND DIRECT CPV \\
  FROM BELLE AND BABAR}

\author{ T.~TOMURA }

\address{Department of Physics, University of Tokyo \\
  7--3--1 Hongo, Bunkyo-ku, Tokyo 113--0033, Japan}

\maketitle\abstracts{
  Recent measurements of branching fractions and $CP$-violating
  partial-rate asymmetries for rare hadronic $B$-meson decays
  by the asymmetric-energy $B$-factory experiments,
  Belle and \babar{}, are reviewed.
}

\section{Introduction}

Recent measurements of the mixing-induced $CP$-violating asymmetry
parameter $\sin2\phi_1$ (or $\sin2\beta$)~\cite{sin2phi1}
strongly supports the Kobayashi-Maskawa (KM) mechanism.~\cite{KM}
However, a full test of the KM mechanism requires additional
measurements for the other angles $\phi_2$ ($\alpha$)
and $\phi_3$ ($\gamma$) of the unitarity triangle.~\cite{UT}
Charmless hadronic decays of $B$ mesons
contain enough information to measure these angles,
but the extraction of unitarity angles from these decay modes
has some difficulty caused by hadronic uncertainties.
However, measurements of enough final states
can provide sufficient constraints to the sizes
of hadronic amplitudes and strong phases,
which are necessary for the extraction of angles.

In the KM scheme, the direct $CP$ violation (DCPV)
is also expected and has already been observed
in the $K$ meson system.~\cite{K_dcpv}
However, this phenomenon has not been observed yet
in the $B$ meson system.  The search for DCPV
is an important issue at $B$-factory experiments.
Charmless hadronic $B$ decays can provide rich sample
for the DCPV search, because many of these decays
are described by $b \to u$ tree and $b \to s$ penguin diagrams.
The interference between the two diagrams can cause
the partial-rate asymmetry \Acp{} as
\begin{align}
  \Acp &\equiv \frac{\Gamma(\bbar \to \overline{f}) - \Gamma(B \to f)}
  {\Gamma(\bbar \to \overline{f}) + \Gamma(B \to f)} \nonumber \\
  &= \frac{2|P||T|\sin\Delta\phi\sin\Delta\delta}
  {|P|^2 + |T|^2 + 2|P||T|\cos\Delta\phi\cos\Delta\delta} ,
\end{align}
where $\Gamma(B \to f)$ denotes the partial width of \bz{} or \bp{}
decaying into a flavor-specific final state $f$
and $\Gamma(\bbar \to \overline{f})$ denotes
that of charge conjugate mode,
$T$ and $P$ represent the tree and penguin amplitudes, respectively,
and $\Delta\phi$ and $\Delta\delta$ are the differences
in weak and strong phases between two amplitudes.
DCPV is also sensitive to the new physics beyond the Standard Model (SM)
through the contribution of new particles to the penguin loop.

In this paper, mainly analyses for $B \to K\pi$, $\pi\pi$, and $KK$
decays, which are referred to as $B \to hh$, are described
and the results for other rare hadronic $B$ decays are summarized.

\section{Analysis}

The analyses of Belle are based on $78~\fb$ data sample collected
at the \UPS{} resonance, which corresponds to $85 \times 10^6$
\bbbar{} pairs, by the Belle detector~\cite{Belle}
at the KEKB $e^+e^-$ storage ring.~\cite{KEKB}
The analyses of \babar{} are based on $81.2~\fb$ data sample
corresponding to $88 \times 10^6$ \bbbar pairs
collected with the \babar{} detector~\cite{BaBar}
at the PEP-II asymmetric-energy $e^+e^-$ collider.~\cite{PEP-II}
The detail of the reconstruction of $B$ mesons and
the event selection is described elsewhere.~\cite{Casey,babar_prl}

Reconstructed $B$ candidates are identified
using two kinematic variables:
the beam-energy constrained mass (or the beam-energy substituted mass)
$\mb ( = \mes ) \equiv \sqrt{(\Ebeam)^2-(\pb)^2}$
and the energy difference $\dE \equiv \Eb - \Ebeam$,
where $\Ebeam$ is the beam energy, $\pb$ and $\Eb$ are
the momentum and energy of the reconstructed $B$ meson
in the center-of-mass system (cms).

The dominant background comes from $e^+e^- \to q\overline{q}$
($q=u$, $d$, $s$, $c$) continuum process.
These backgrounds are suppressed by the event topology.
Belle uses the likelihood ratio calculated from two variables:
the modified Fox-Wolfram moments~\cite{FW,Casey}
that are combined using a Fisher discriminant into a single variable
and the angle of the $B$ flight direction with respect to the beam axis.
\babar{} uses the angle between the sphericity axis
of the $B$ candidate and the sphericity axis
of the remaining particles in that event
and the Fisher discriminant calculated from the momenta
of remaining particles and the angles between their momenta
and the thrust axis of $B$ candidate in the cms.

For the final states that include a charged pion or kaon,
high momentum particle identification (PID) is important.
PID of Belle is based on the light yield
in the aerogel Cherenkov counter (ACC)
and \dedx{} measurements in the central drift chamber (CDC).
PID of \babar{} is accomplished with the Cherenkov angle measurement
from a detector of internally reflected Cherenkov light (DIRC).

\section{Result}

\subsection{Branching Fraction}

\begin{figure}
  \begin{center}
    \epsfig{figure=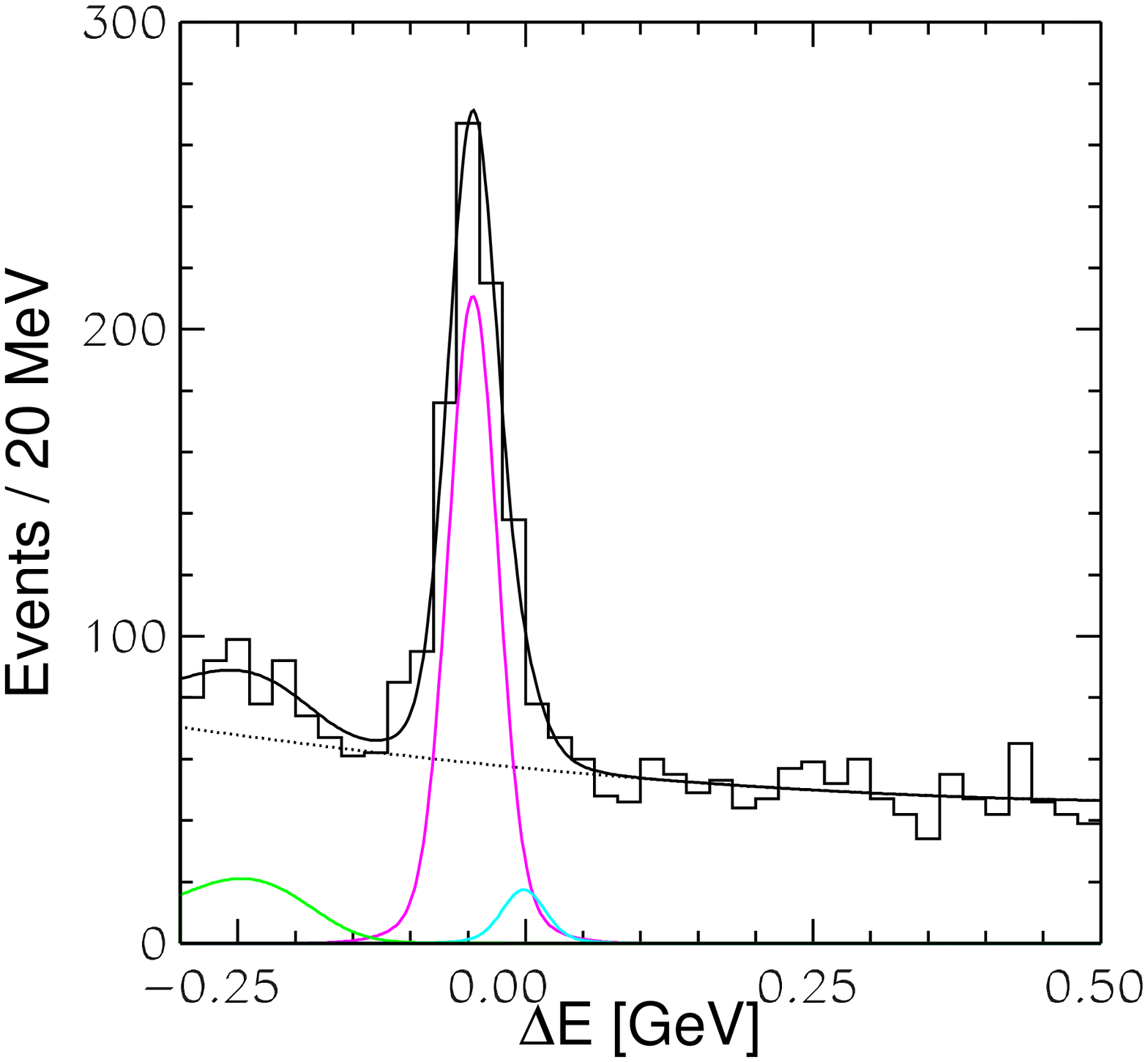,width=0.32\textwidth}
    \epsfig{figure=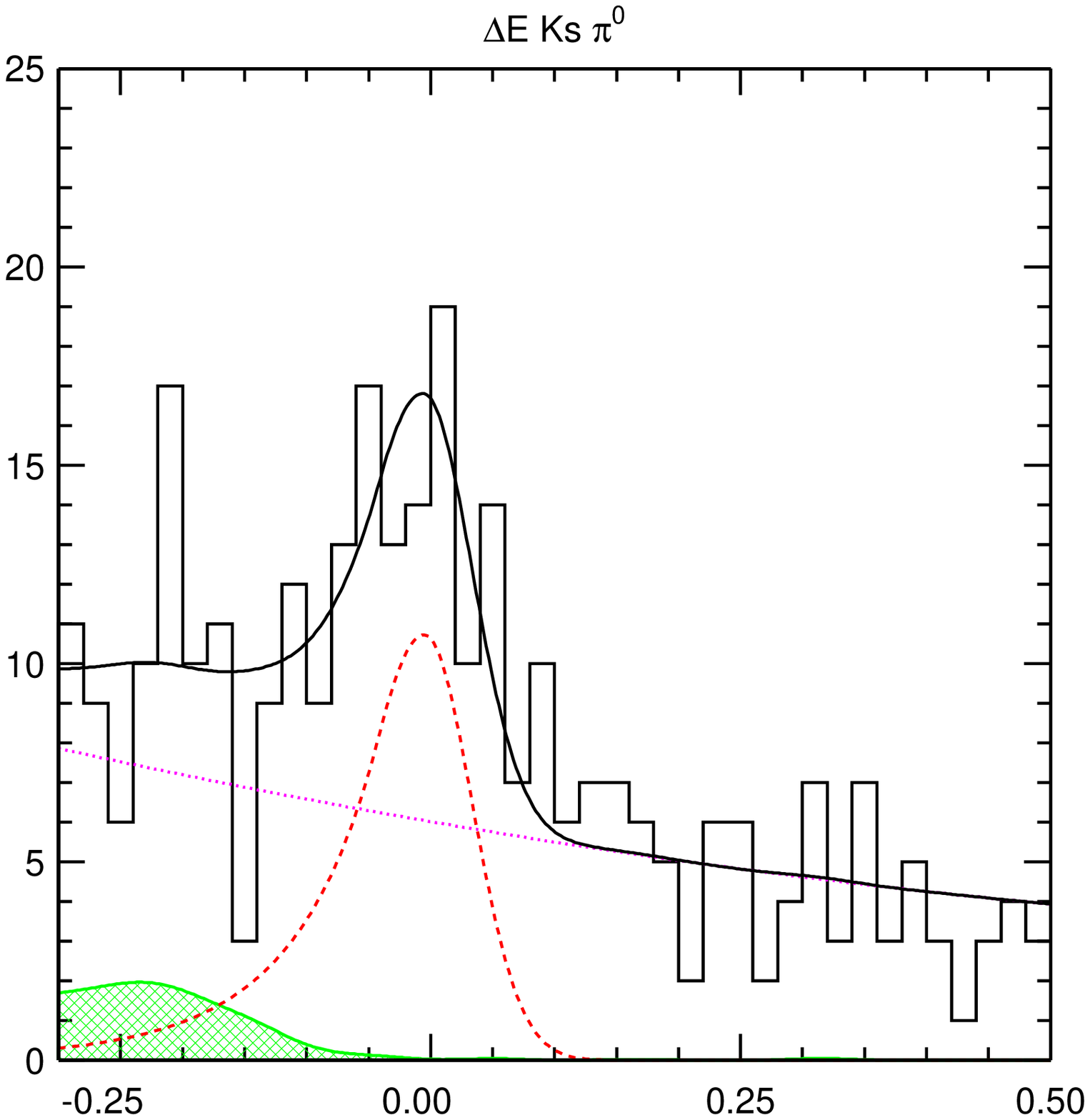,width=0.32\textwidth}
    \epsfig{figure=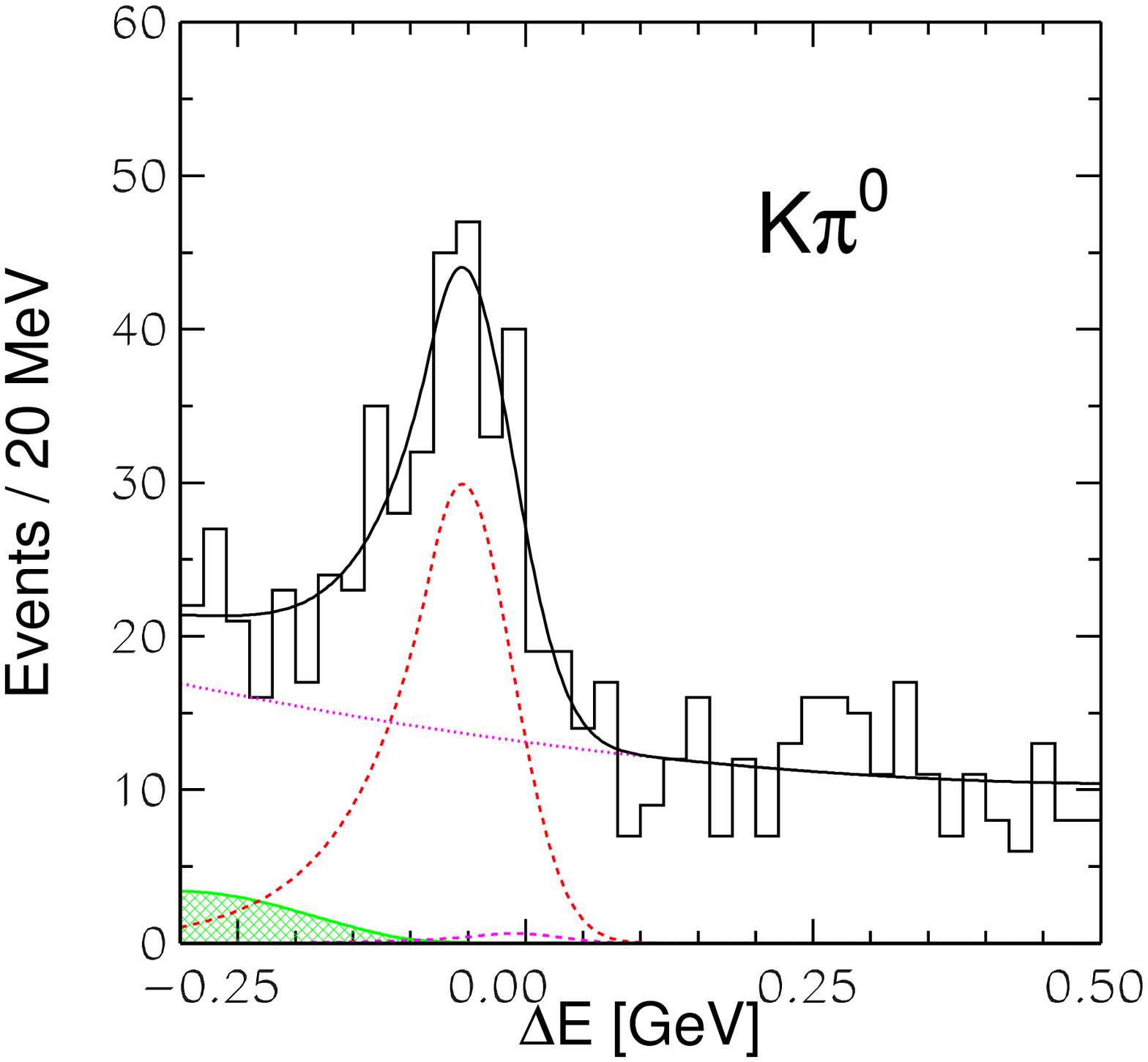,width=0.32\textwidth} \\
    \vspace*{-1ex}
    \makebox[0.32\textwidth][l]{\hspace*{2em} (a) $\bz \to \kp\pim$}
    \makebox[0.32\textwidth][l]{\hspace*{2em} (b) $\bz \to \ks\piz$}
    \makebox[0.32\textwidth][l]{\hspace*{2em} (c) $\bp \to \kp\piz$} \\
    \epsfig{figure=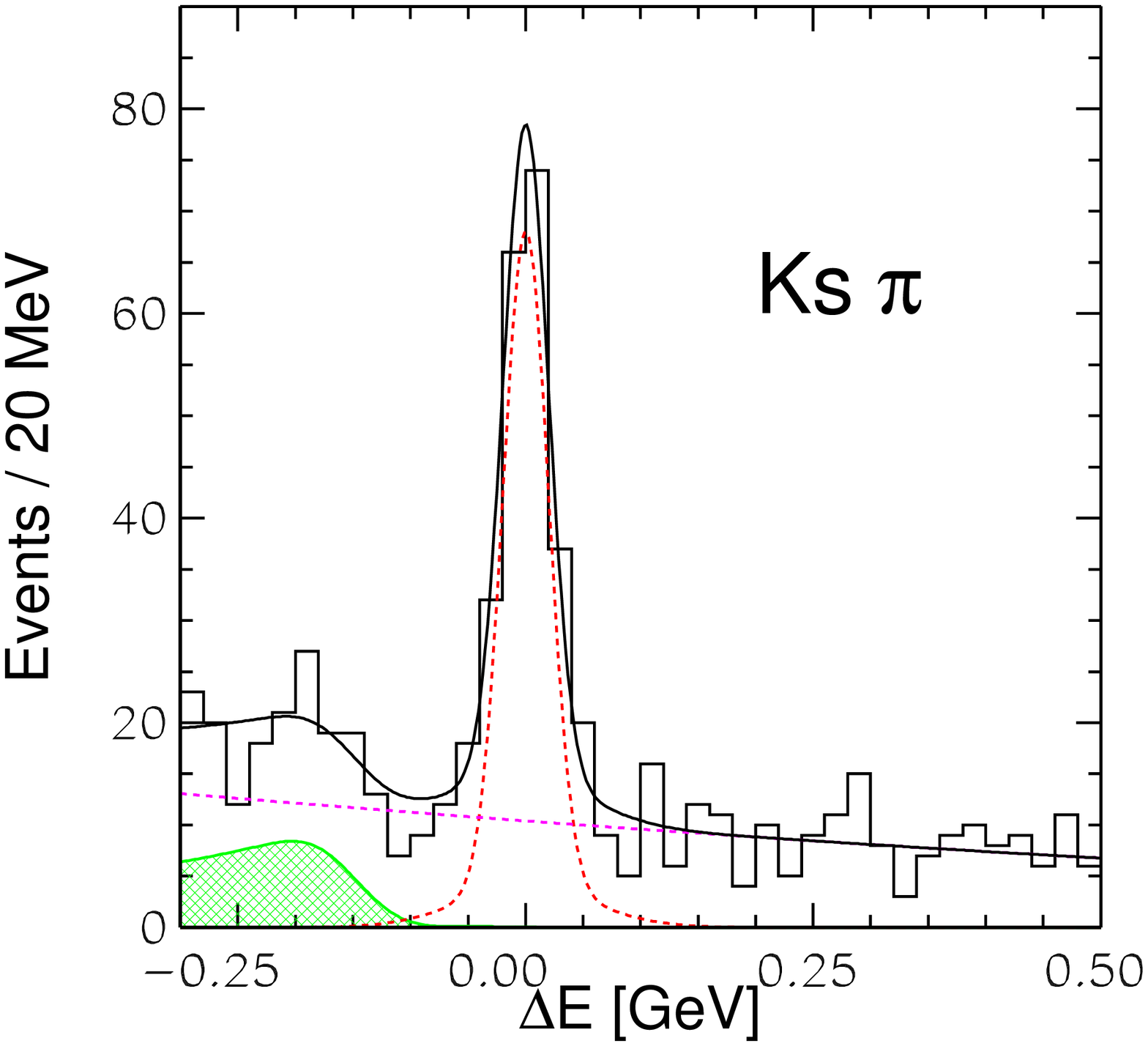,width=0.32\textwidth}
    \epsfig{figure=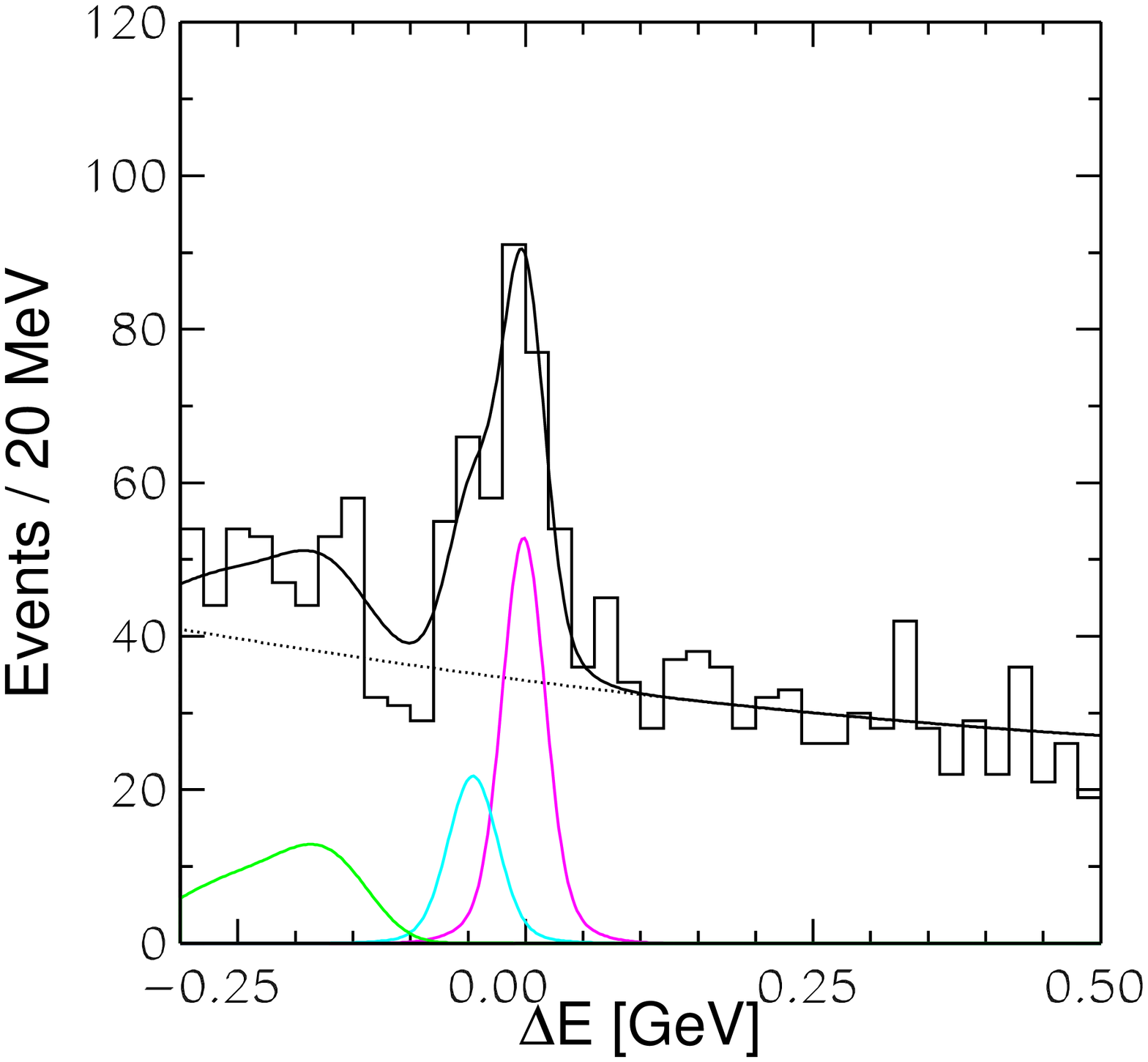,width=0.32\textwidth}
    \epsfig{figure=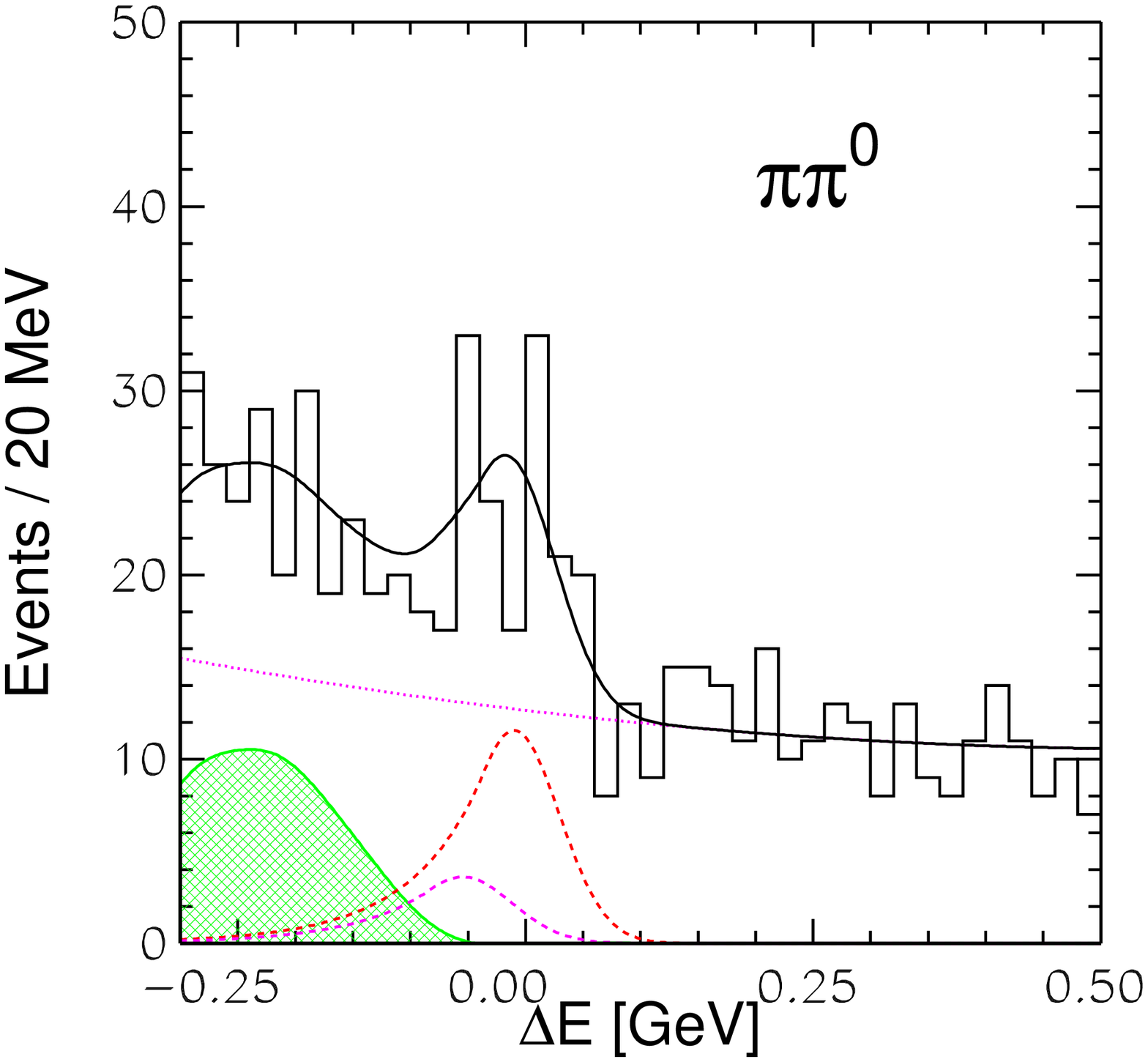,width=0.32\textwidth} \\
    \vspace*{-1ex}
    \makebox[0.32\textwidth][l]{\hspace*{2em} (d) $\bp \to \ks\pip$}
    \makebox[0.32\textwidth][l]{\hspace*{2em} (e) $\bz \to \pip\pim$}
    \makebox[0.32\textwidth][l]{\hspace*{2em} (f) $\bp \to \pip\piz$} \\
    \epsfig{figure=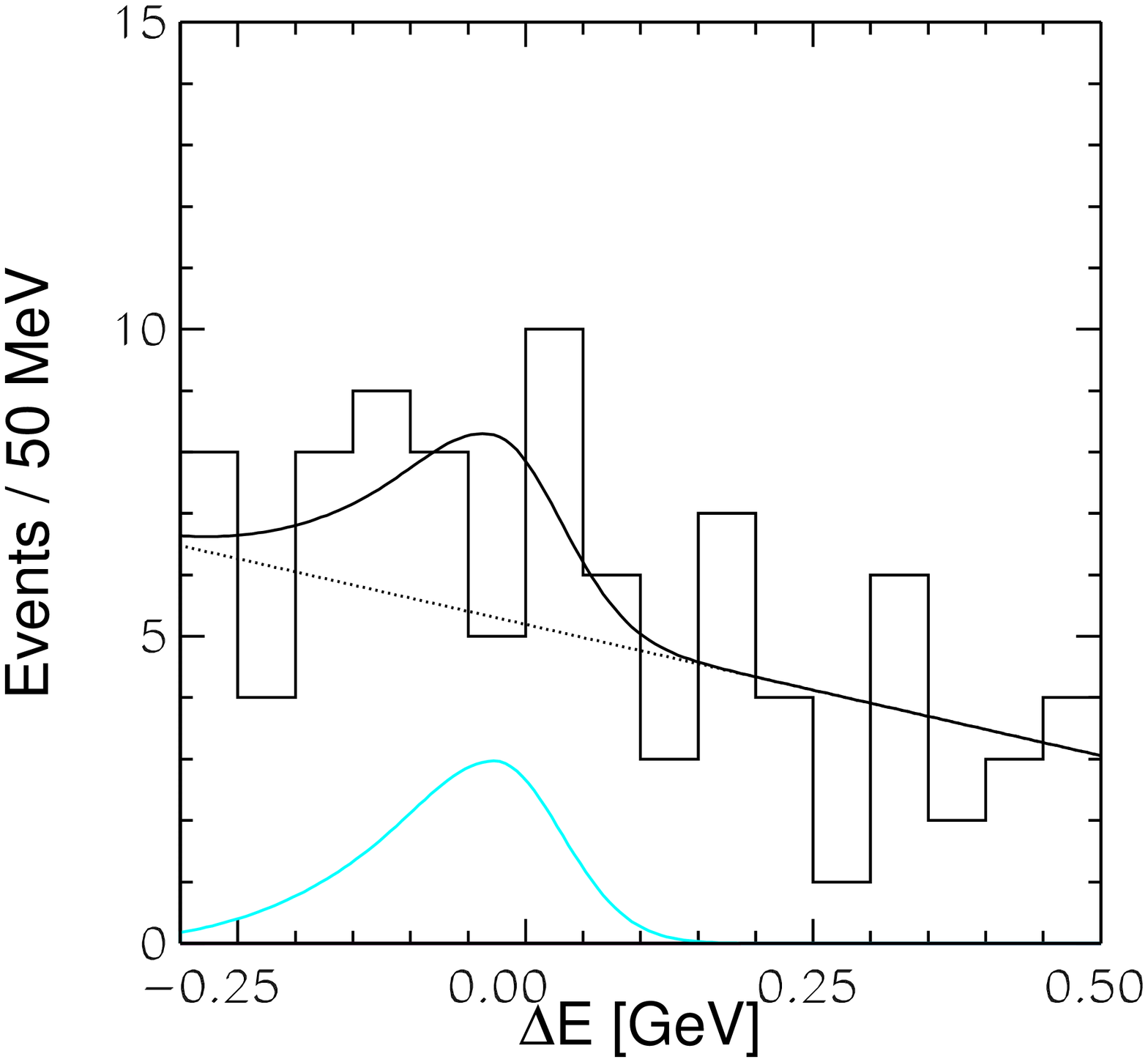,width=0.32\textwidth}
    \epsfig{figure=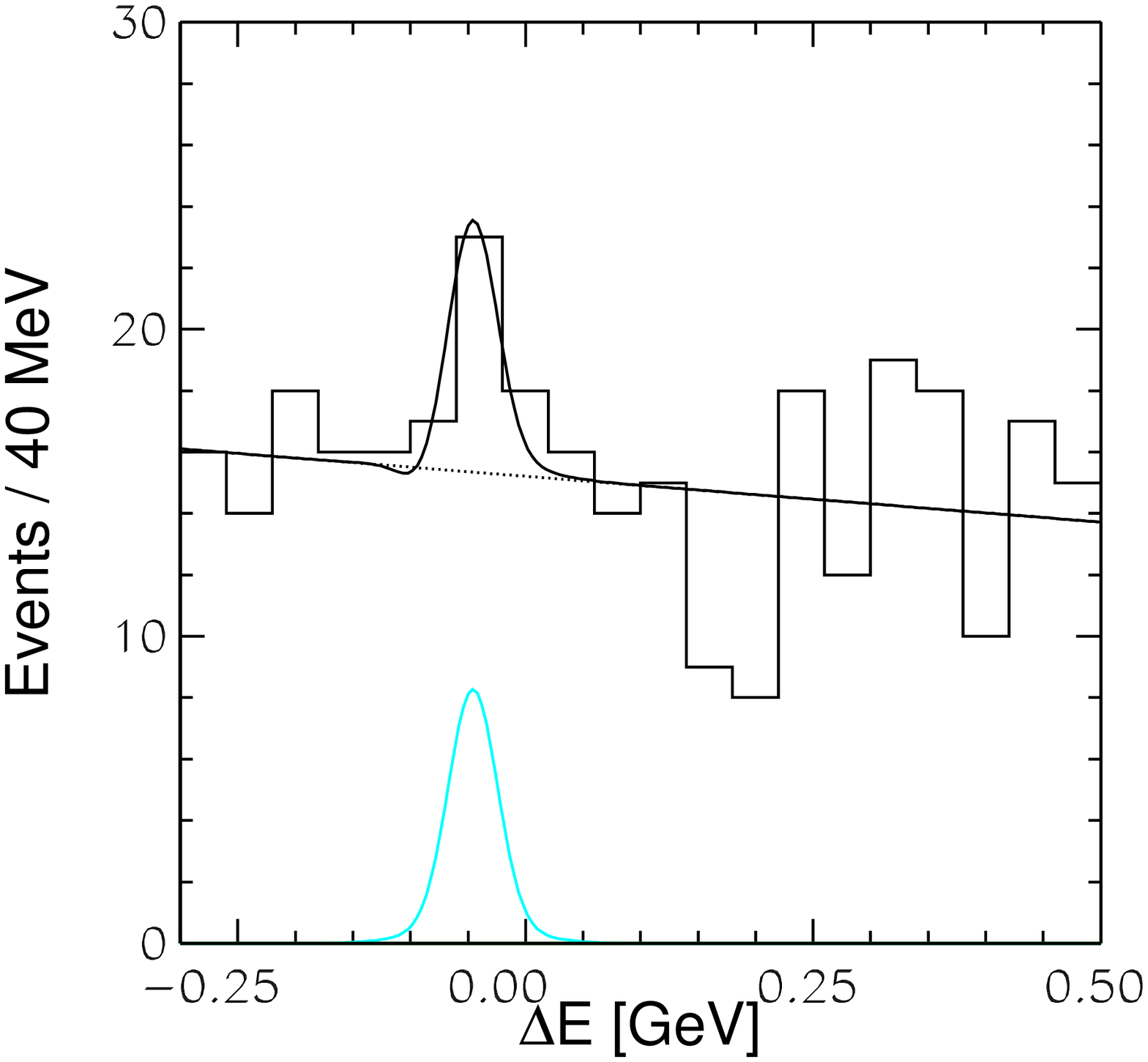,width=0.32\textwidth}
    \epsfig{figure=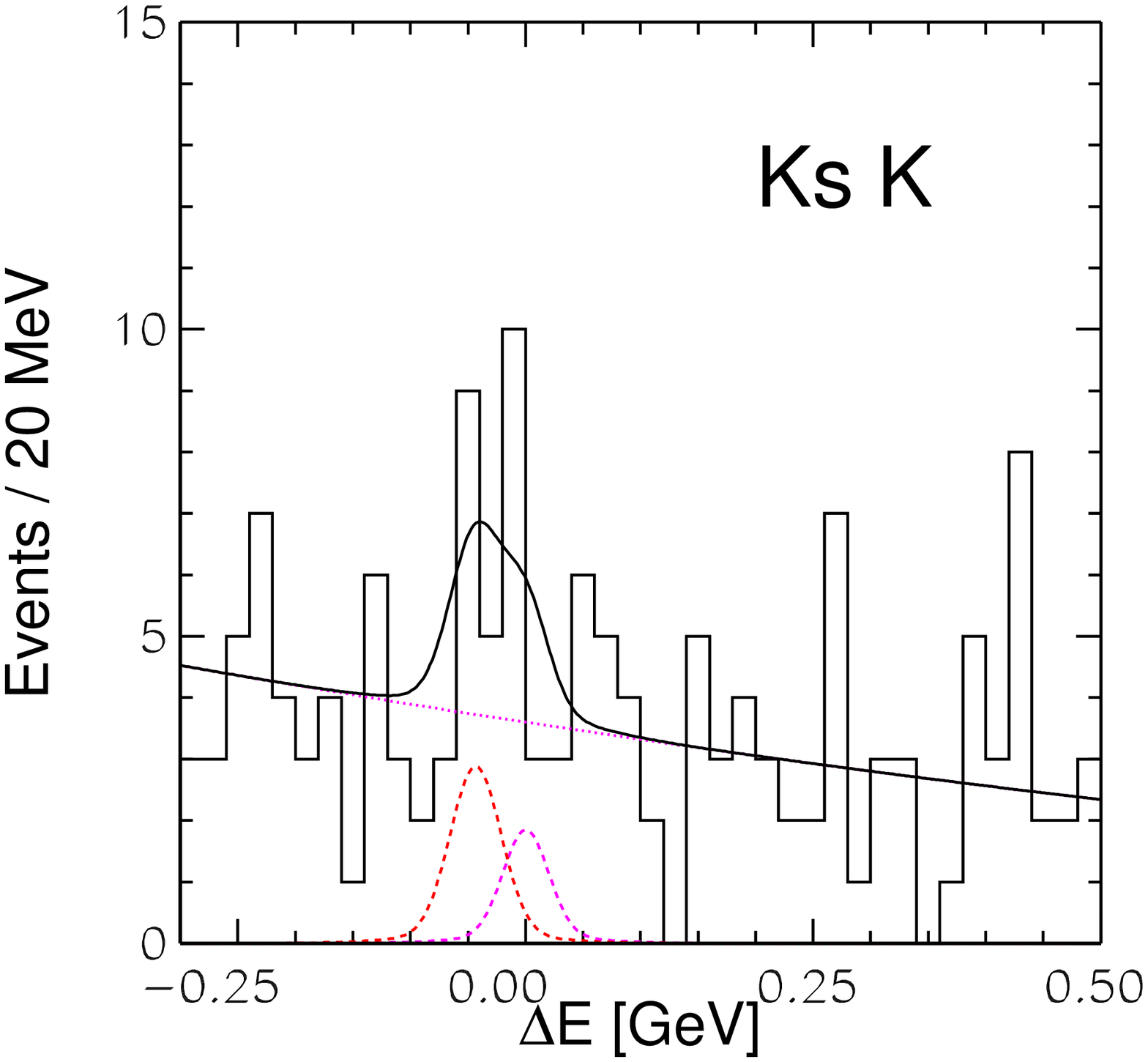,width=0.32\textwidth} \\
    \vspace*{-1ex}
    \makebox[0.32\textwidth][l]{\hspace*{2em} (g) $\bz \to \piz\piz$}
    \makebox[0.32\textwidth][l]{\hspace*{2em} (h) $\bz \to \kp\km$}
    \makebox[0.32\textwidth][l]{\hspace*{2em} (i) $\bp \to \kp\ks$}
    \epsfig{figure=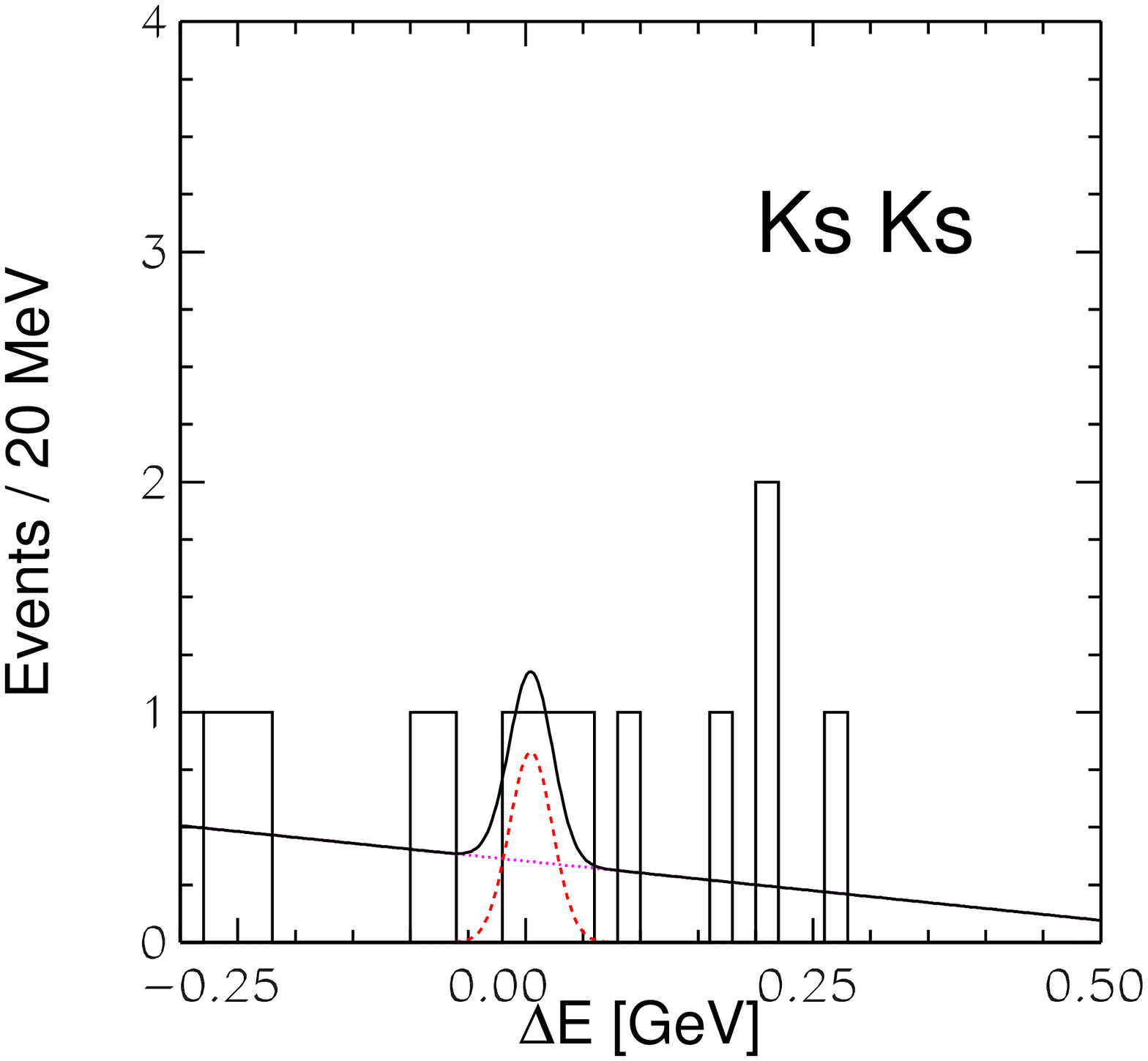,width=0.32\textwidth} \\
    \vspace*{-1ex}
    \makebox[0.32\textwidth][l]{\hspace*{2em} (j) $\bz \to \ks\ks$}
  \end{center}
  \caption{Distributions of \dE{} and fit results for $B \to hh$ modes
    from Belle.  Results of the fits are also shown. \label{fig:belle_hh}}
\end{figure}

Figure~\ref{fig:belle_hh} shows the \dE{} distributions
obtained by the Belle experiment for $B \to hh$ modes
in the \mb{} signal region.
The signal yields are extracted by a binned maximum likelihood fit
to the \dE{} distribution.
The \dE{} fits include four components:
signal, crossfeed from other misidentified signals,
continuum background, and backgrounds from multibody and radiative
charmless $B$ decays.  The results of the fits are
also shown in Fig.~\ref{fig:belle_hh}.

\begin{figure}
  \begin{center}
    \epsfig{figure=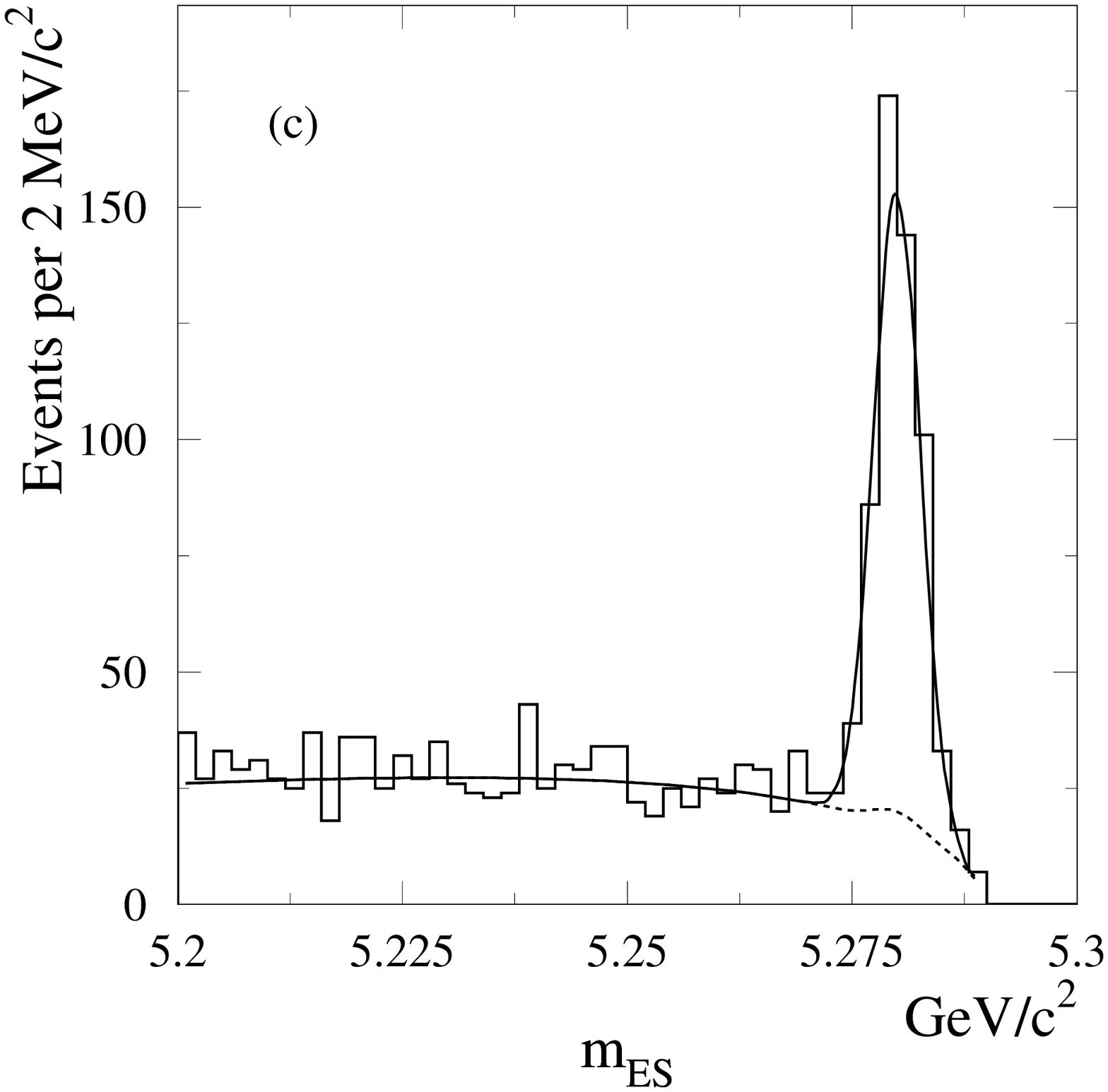,width=0.32\textwidth}
    \epsfig{figure=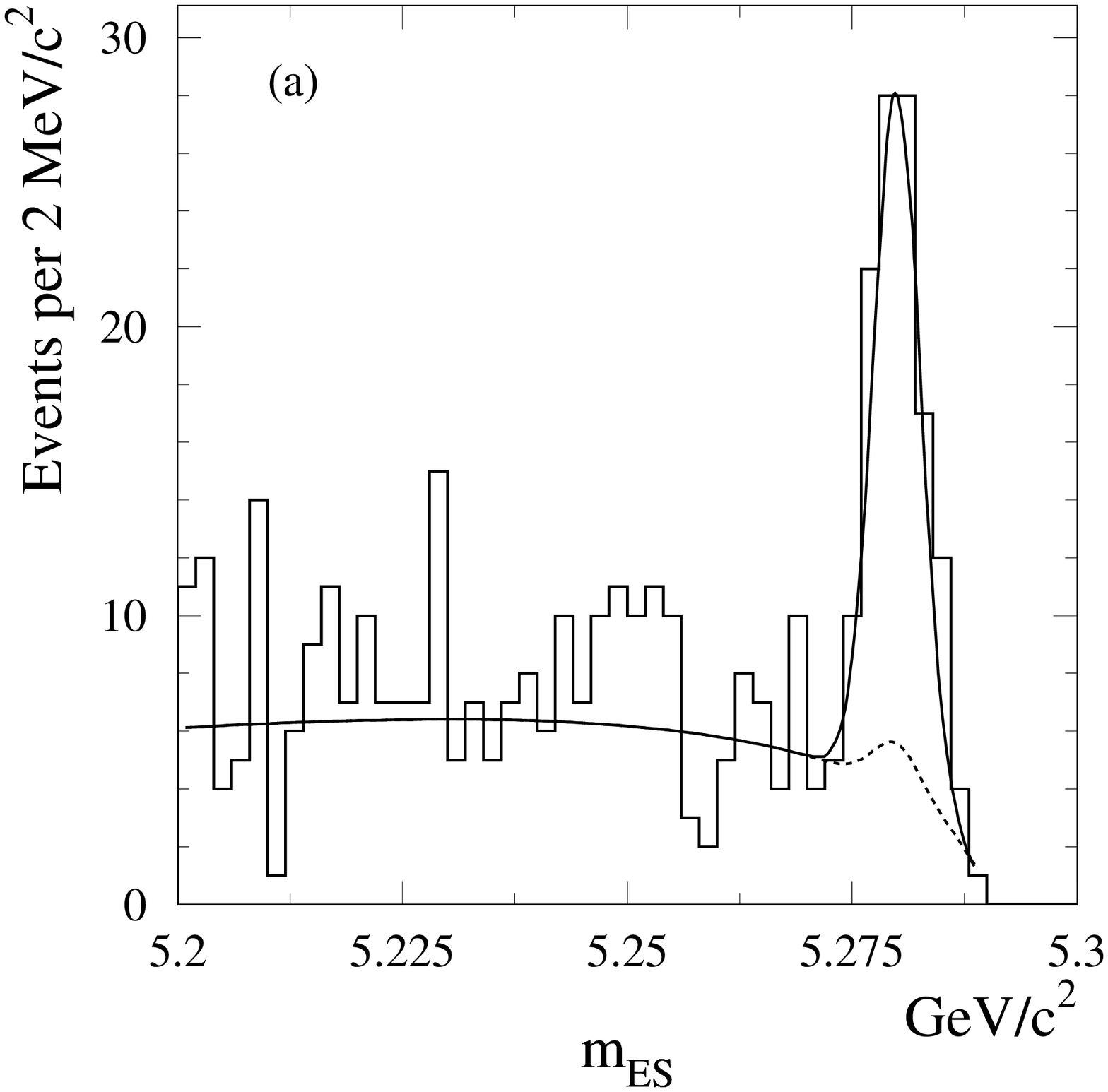,width=0.32\textwidth}
    \epsfig{figure=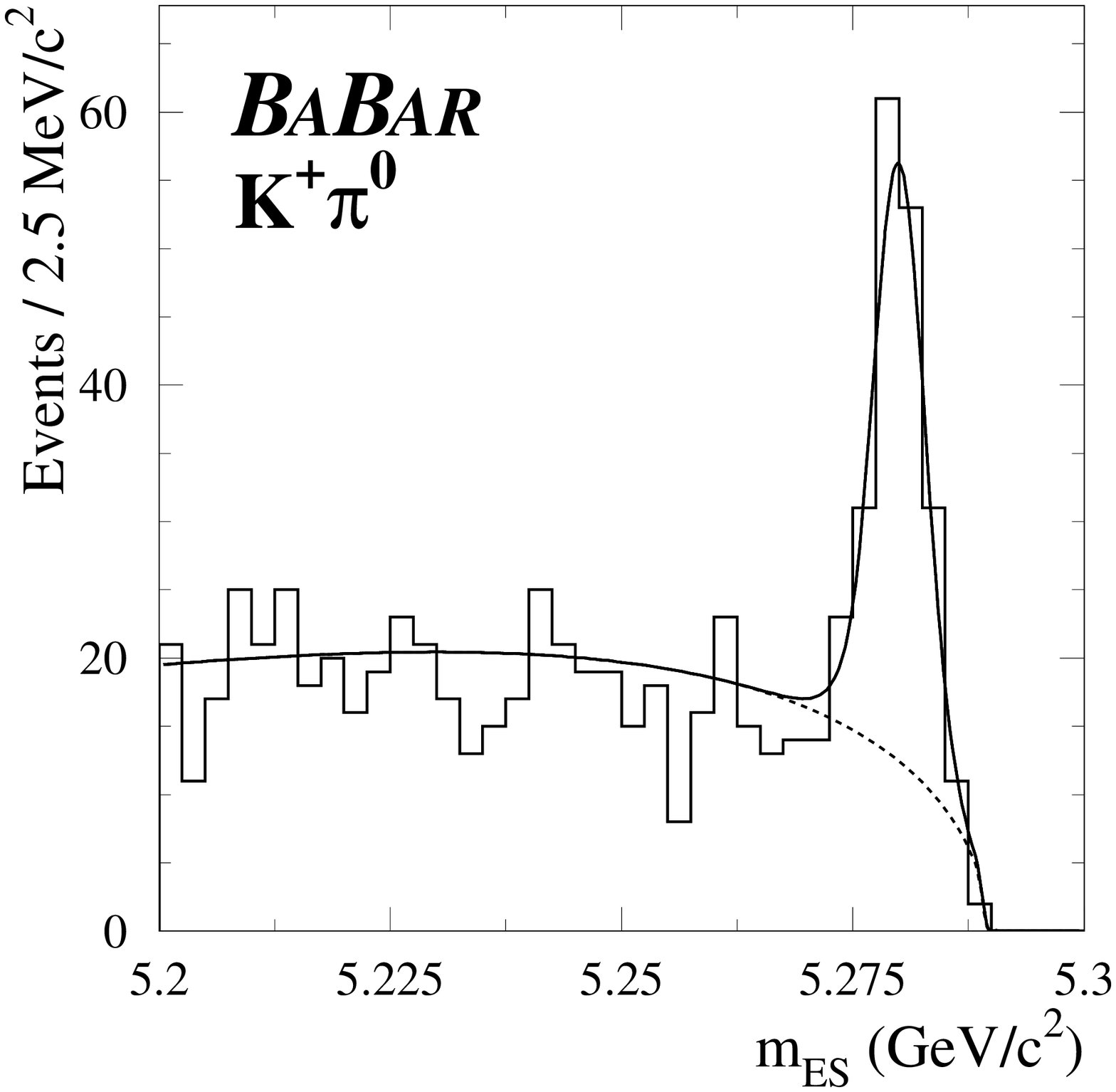,width=0.32\textwidth} \\
    \epsfig{figure=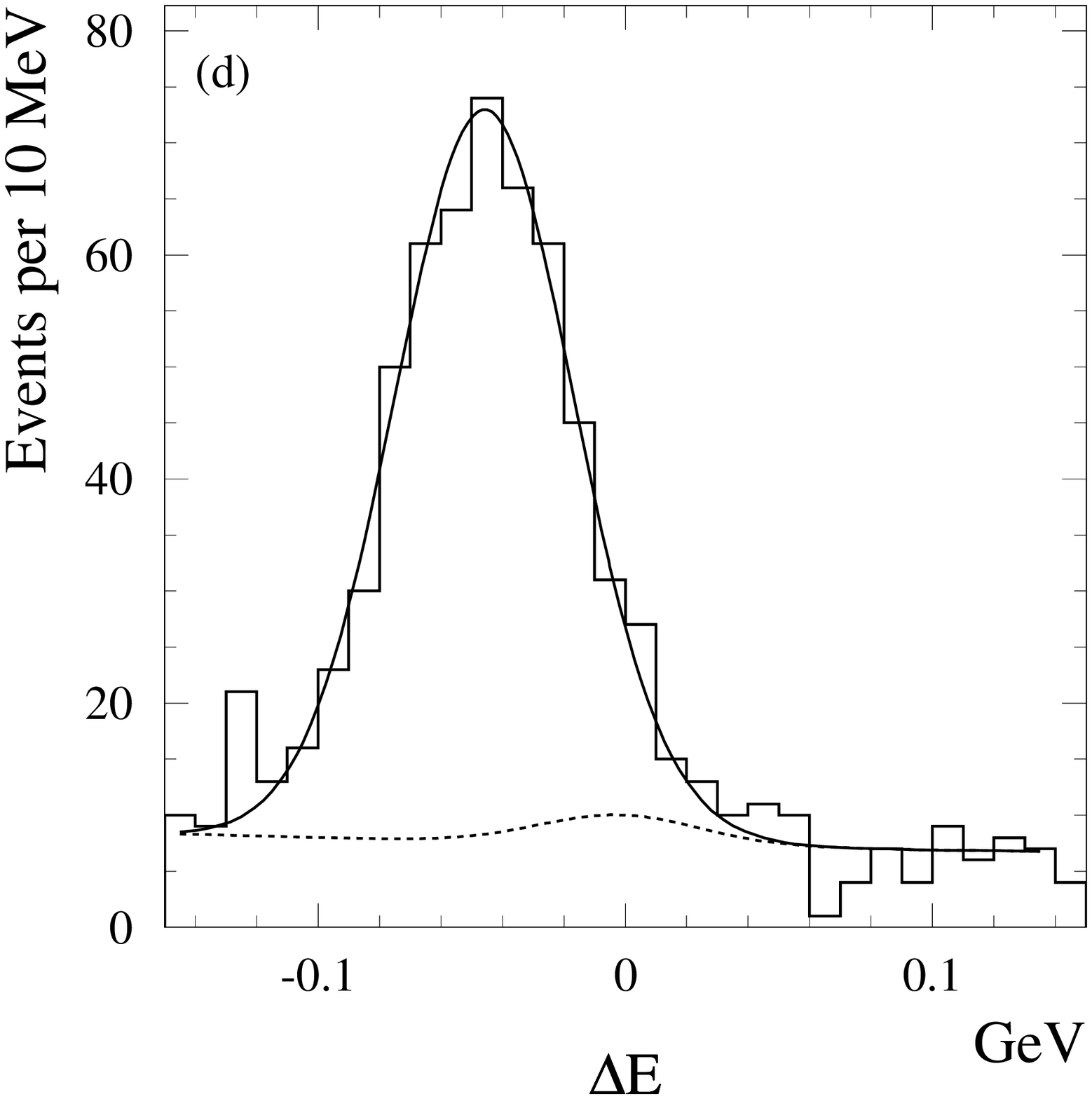,width=0.32\textwidth}
    \epsfig{figure=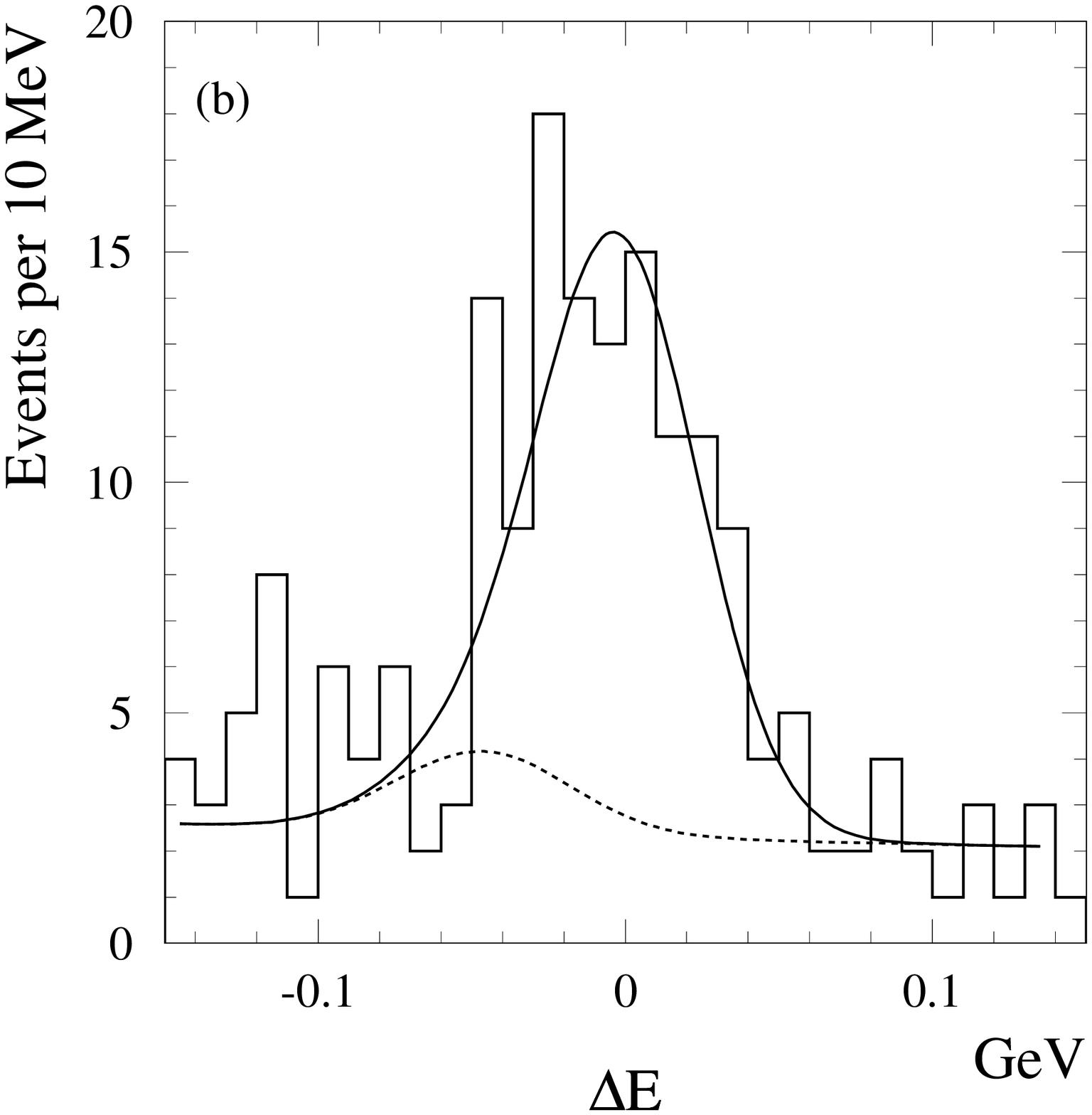,width=0.32\textwidth}
    \epsfig{figure=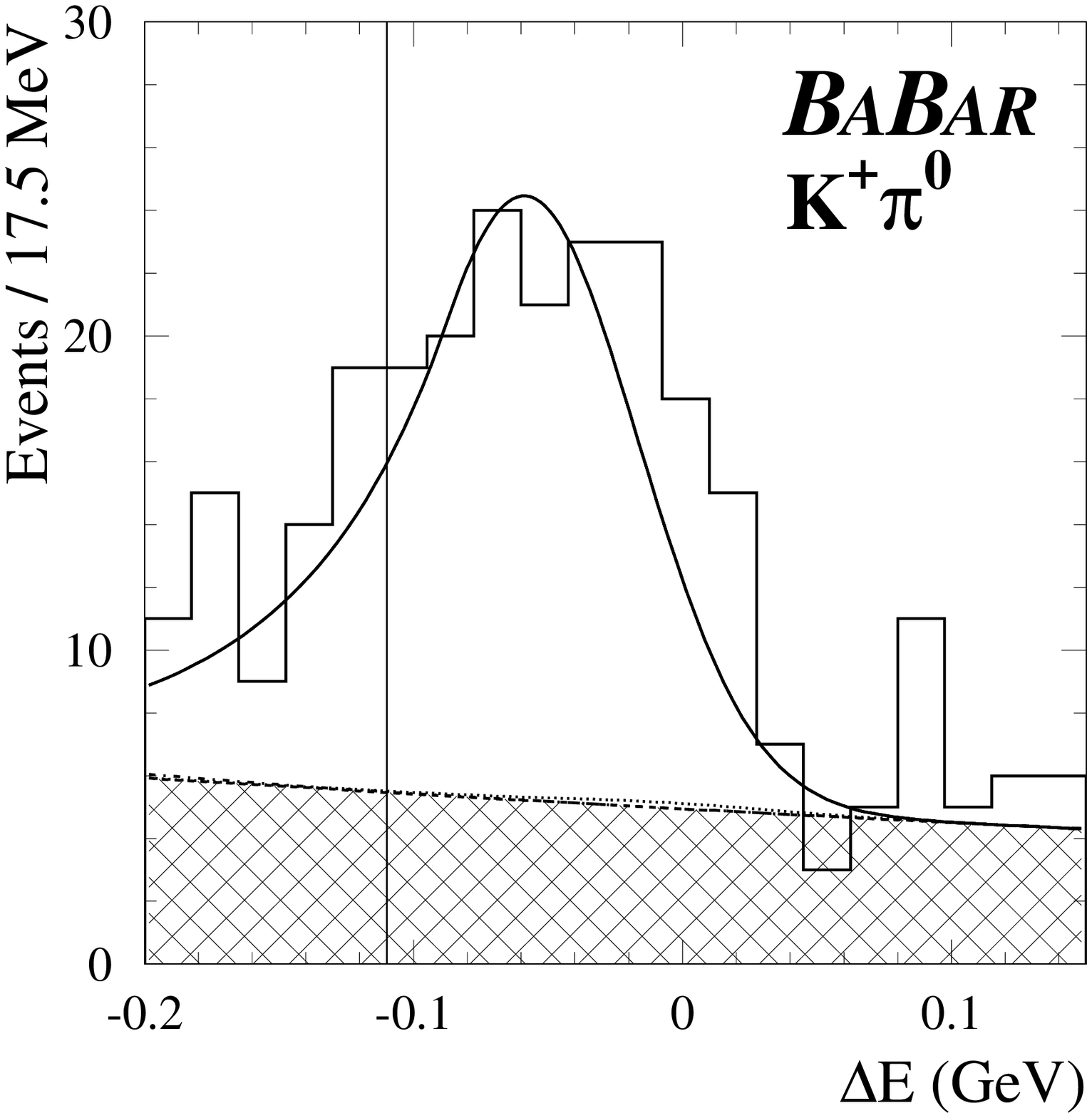,width=0.32\textwidth} \\
    \makebox[0.32\textwidth][l]{\hspace*{2em} (a) $\bz \to \kp\pim$}
    \makebox[0.32\textwidth][l]{\hspace*{2em} (b) $\bz \to \pip\pim$}
    \makebox[0.32\textwidth][l]{\hspace*{2em} (c) $\bp \to \kp\piz$} \\
    \vspace*{3ex}
    \epsfig{figure=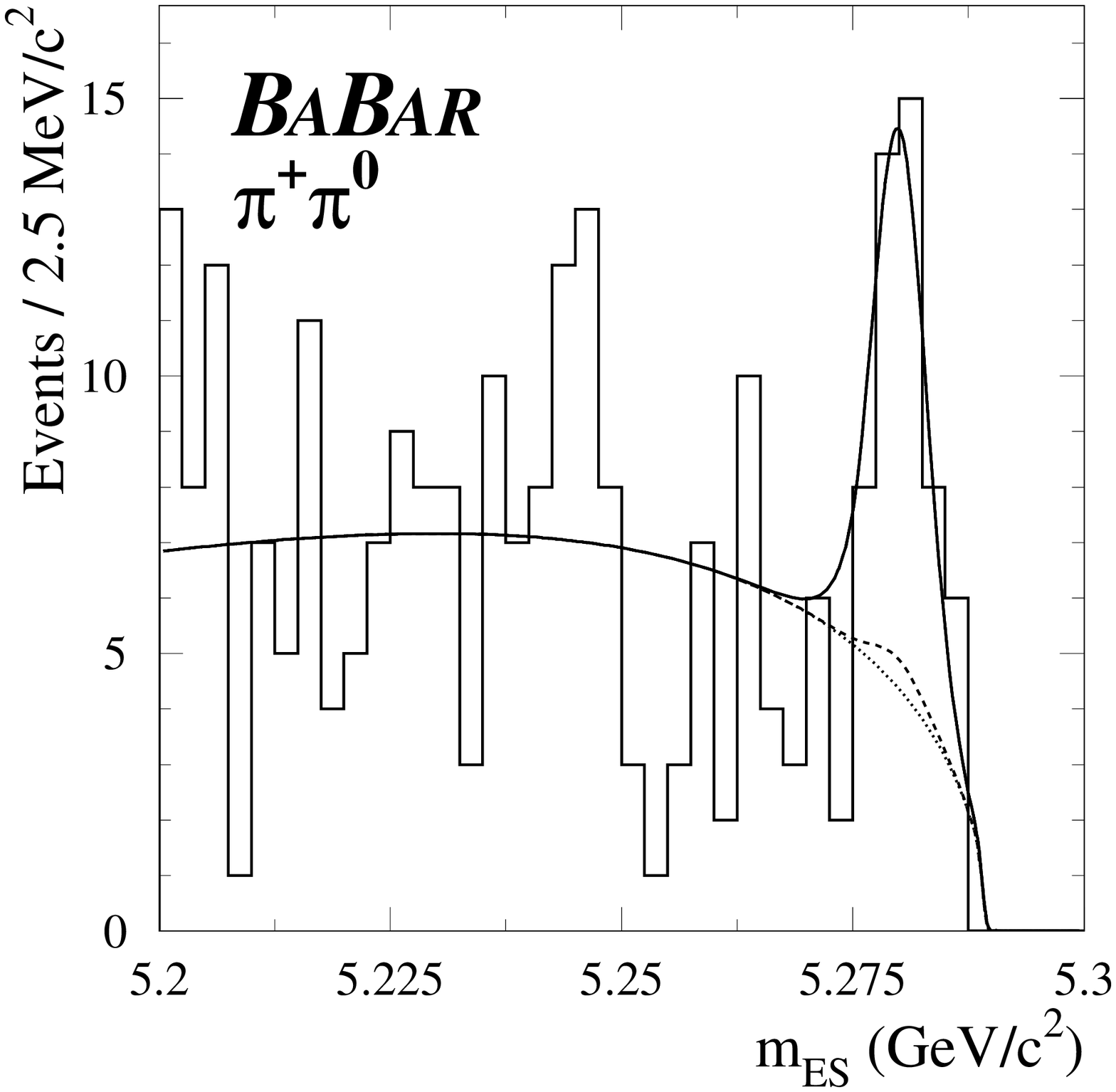,width=0.32\textwidth}
    \epsfig{figure=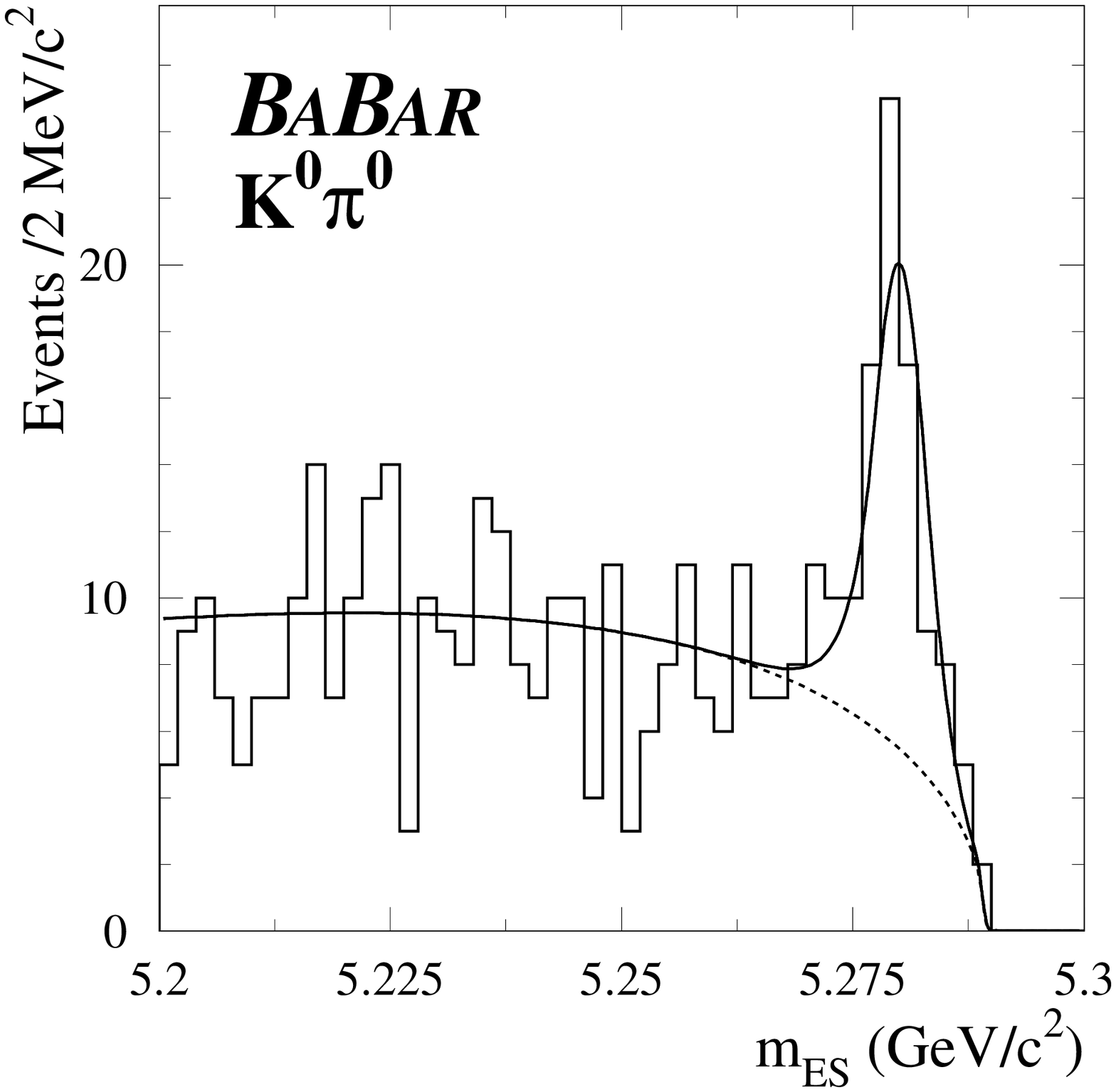,width=0.32\textwidth}
    \epsfig{figure=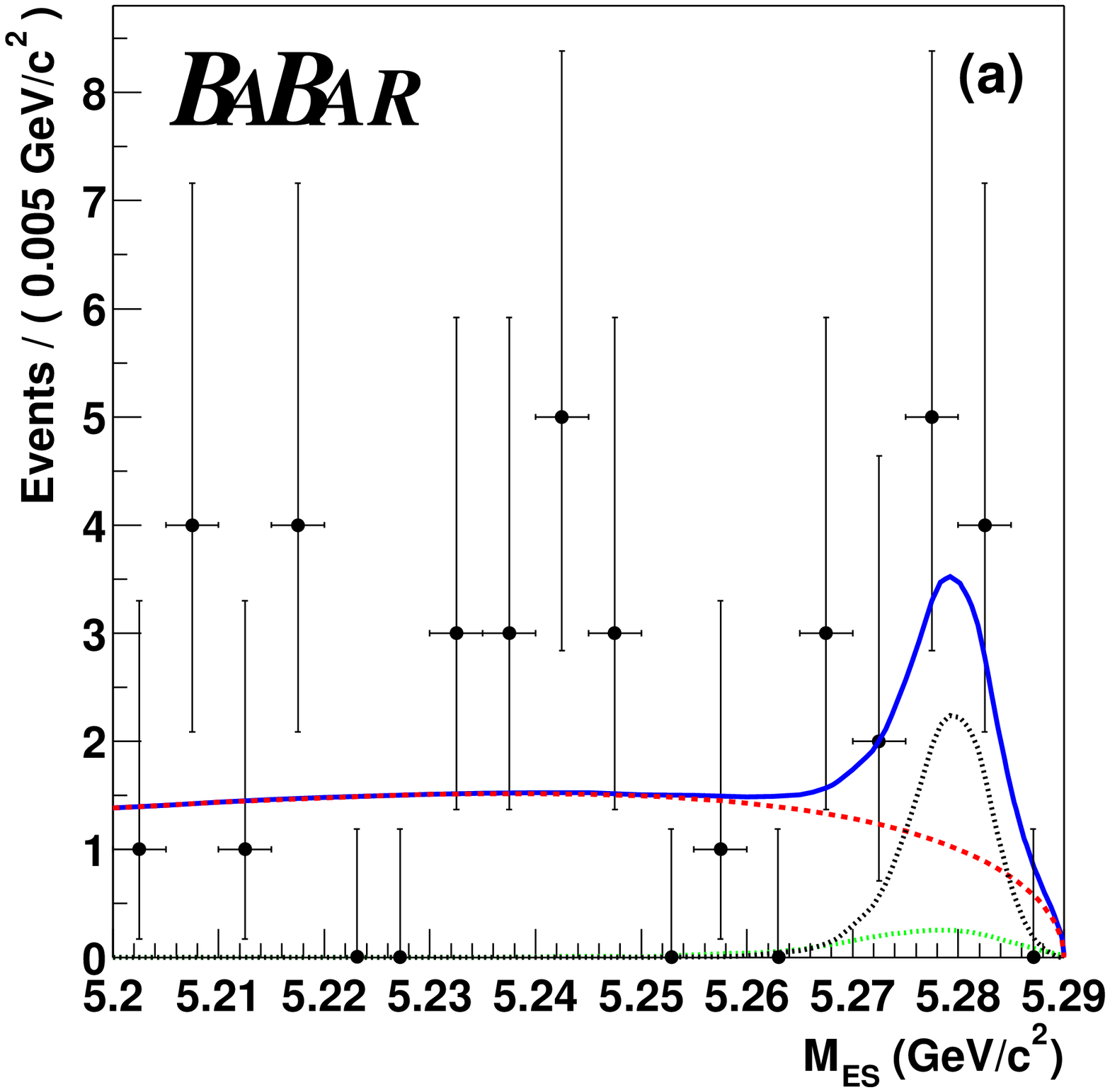,width=0.32\textwidth} \\
    \epsfig{figure=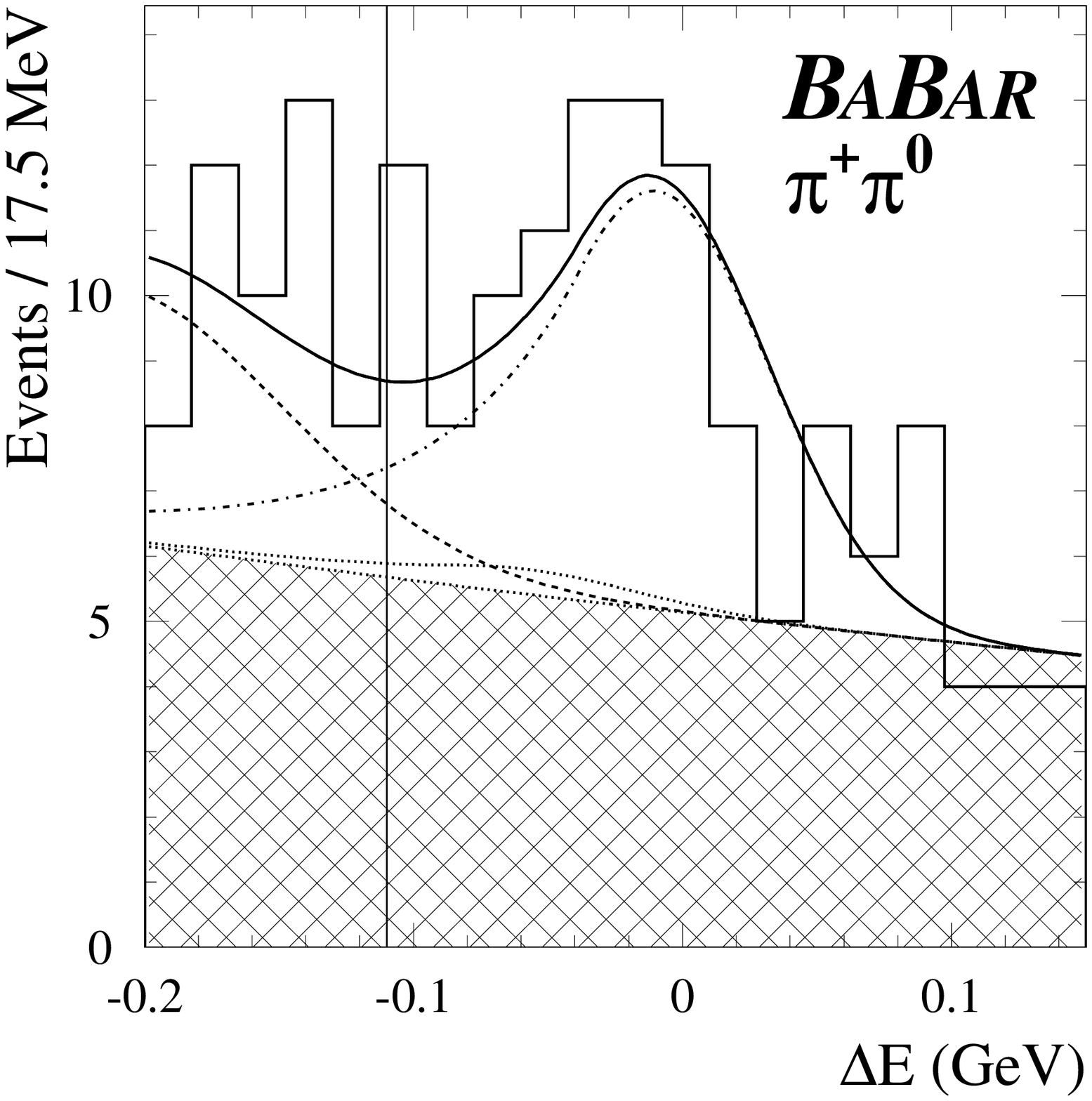,width=0.32\textwidth}
    \epsfig{figure=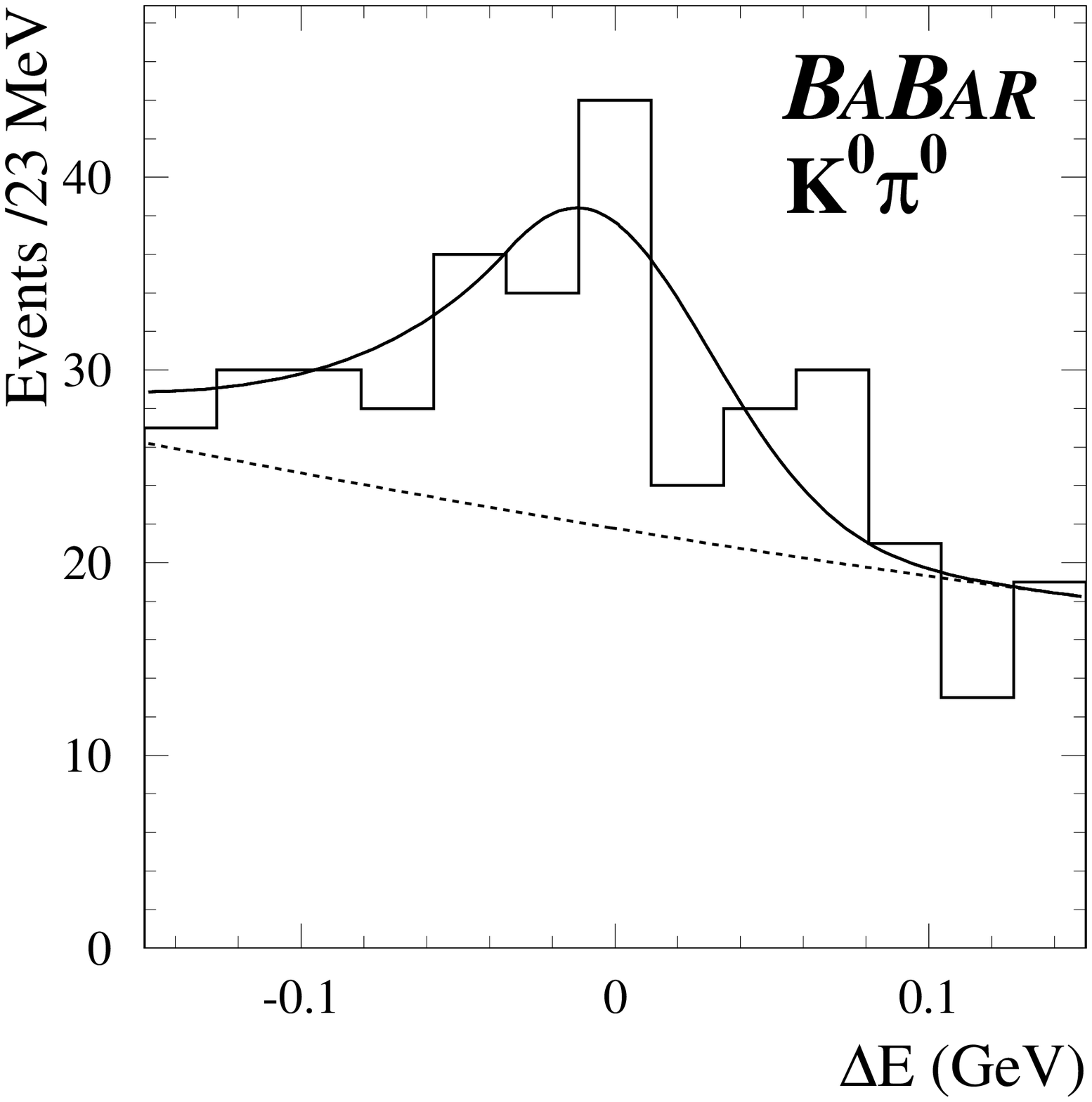,width=0.32\textwidth}
    \epsfig{figure=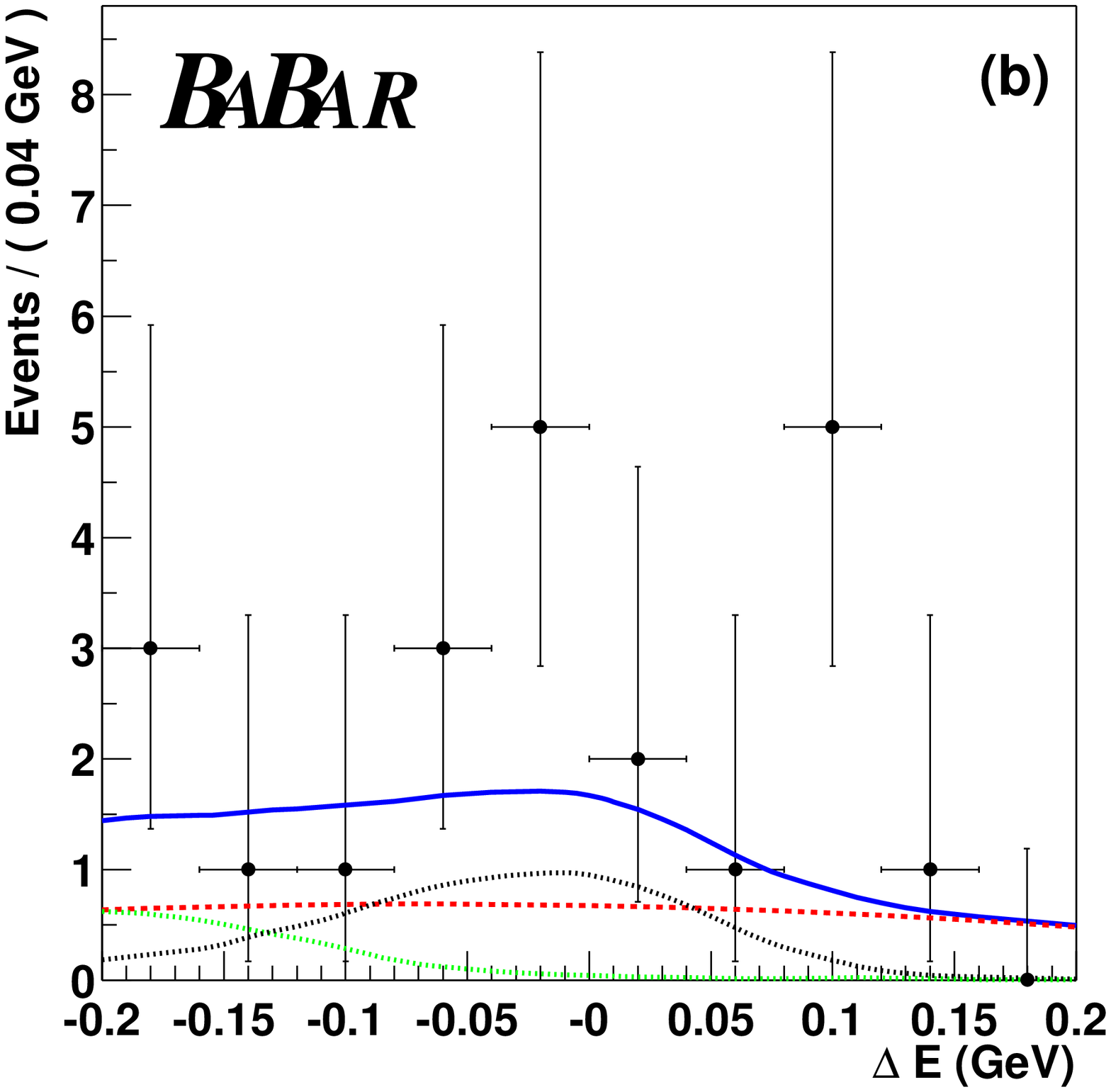,width=0.32\textwidth} \\
    \makebox[0.32\textwidth][l]{\hspace*{2em} (d) $\bp \to \pip\piz$}
    \makebox[0.32\textwidth][l]{\hspace*{2em} (e) $\bz \to \ks\piz$}
    \makebox[0.32\textwidth][l]{\hspace*{2em} (f) $\bp \to \piz\piz$}
  \end{center}
  \caption{Distributions of \mes{} and \dE{} for $B \to hh$ modes
    from \babar{}.  Results of the fits are also shown. \label{fig:babar_hh}}
\end{figure}

Figure~\ref{fig:babar_hh} shows the distributions of \mes{}
and \dE{} obtained by the \babar{} experiment
after the selection to enhance the signal purity.~\cite{babar_prl,babar_ichep02}
\babar{} uses an unbinned extended maximum likelihood fit
to extract the signal yields.
The input variables to the fit are \mes{}, \dE{},
Fisher discriminant of the event shape parameter,
Cherenkov angles for the charged tracks.
The projections of the fit results are also shown in Fig.~\ref{fig:babar_hh}.

Using the signal yields obtained from the fit and
the reconstruction efficiencies,
the branching fractions are derived and listed in Table~\ref{tab:bf_kpi}.
\begin{table}[tbp]
  \caption{Branching fractions \Br{} for $B \to hh$ modes
    obtained from Belle and \babar{}.
    \babar{} uses $54~\fb$ data sample for $\bp \to \kz\pip$
    and $\kp\kzb$ modes. \label{tab:bf_kpi}}
  \vspace{0.4cm}
  \begin{center}
    \newcommand{\m}{\hphantom{$-$}}
    \newcommand{\hdot}{\hphantom{.}}
    \newcommand{\hdig}{\hphantom{0}}
    \newcommand{\hdd}{\hdot\hdig}
    \newcommand{\cc}[1]{\multicolumn{1}{c|}{#1}}
    \newcommand{\ccc}[1]{\multicolumn{1}{|c|}{#1}}
    \newcommand{\hsa}{\hspace{-0.7mm}}
    \newcommand{\hsb}{\hspace{+1.6mm}}
    \newcommand{\hsd}{\hspace{+1.1mm}}
    \newcommand{\hse}{\hspace{+2.0mm}}
    \newcommand{\hsf}{\hspace{+0.2mm}}
    \newcommand{\aer}[2]{\mbox{$^{\hsd+\hse #1}_{\hsd-\hse #2}$\hsf}}
    \begin{tabular}{|l|l|l|}
      \hline
      \ccc{Mode} & \cc{$\Br$ [$10^{-6}$] (\babar{})}
      & \cc{$\Br$ [$10^{-6}$] (Belle)} \\
      \hline
      $\bz \to \kp\pim$
      & $17.9 \pm 0.9 \pm 0.7$
      & $18.5\pm 1.0\pm 0.7$ \\
      $\bp \hsa \to \kp\piz$
      & $12.8 \aer{1.2}{1.1} \pm 1.0$
      & $12.8\pm 1.4\aer{1.4}{1.0}$ \\
      $\bp \hsa \to \kz\pip$
      & $17.5 \aer{1.8}{1.7} \pm 1.3$
      & $22.0\pm 1.9\pm 1.1$ \\
      $\bz \to \kz\piz$
      & $10.4 \pm 1.5 \pm 0.8$
      & $12.6\pm 2.4\pm 1.4$ \\
      $\bz \to \pip\pim$
      & $\hdig 4.7 \pm 0.6 \pm 0.2$
      & $\hdig 4.4\pm 0.6\pm 0.3$ \\
      $\bp \hsa \to \pip\piz$
      & $\hdig 5.5 \aer{1.0}{0.9} \pm 0.6$
      & $\hdig 5.3\pm 1.3\pm 0.5$ \\
      $\bz \to \piz\piz$
      & $\hdig 1.6 \aer{0.7}{0.6} \aer{0.6}{0.3} < 3.6$
      & $\hdig 1.8\aer{1.4}{1.3}\aer{0.5}{0.7} < 4.4$ \\
      $\bz \to \kp\km$
      & \cc{$< 0.6$}
      & \cc{$< 0.7$} \\
      $\bp \hsa \to \kp\kzb$
      & $-0.6 \aer{0.6}{0.7} \pm 0.3 < 1.3$
      & $\hdig 1.7 \pm 1.2 \pm 0.1 < 3.4$ \\
      $\bz \to \kz\kzb$
      & \cc{---}
      & $\hdig 0.8 \pm 0.8 \pm 0.1 < 3.2$ \\
      \hline
    \end{tabular}
  \end{center}
\end{table}
The branching fractions for other hadronic rare decay modes
are reported by Belle and \babar{} as listed
in Table~\ref{tab:bf_other}.~\cite{rare_hfag}
\begin{table}[tbp]
  \caption{Branching fractions \Br{} for rare hadronic $B$ decays
    other than $B \to hh$ modes.  \label{tab:bf_other}}
  \vspace{0.4cm}
  \begin{center}
    \newcommand{\m}{\hphantom{$-$}}
    \newcommand{\hdot}{\hphantom{.}}
    \newcommand{\hdig}{\hphantom{0}}
    \newcommand{\hdd}{\hdot\hdig}
    \newcommand{\hb}{\hphantom{\mbox{$B^+\to$}}}
    \newcommand{\cc}[1]{\multicolumn{1}{c|}{#1}}
    \newcommand{\ccc}[1]{\multicolumn{1}{|c|}{#1}}
    \newcommand{\hsa}{\hspace{-0.7mm}}
    \newcommand{\hsb}{\hspace{+5.0mm}}
    \newcommand{\hsd}{\hspace{+1.1mm}}
    \newcommand{\hse}{\hspace{+1.6mm}}
    \newcommand{\aer}[2]{\mbox{$^{\hsd+\hse #1}_{\hsd-\hse #2}$}}
    \newlength{\tmp}
    \settowidth{\tmp}{$\bz \to {}$}
    \newcommand{\hsg}{\hspace{\tmp}}
    \begin{tabular}{|l|l|l|}
      \hline
      \ccc{Mode} & \cc{$\Br$ [$10^{-6}$] (\babar{})}
      & \cc{$\Br$ [$10^{-6}$] (Belle)} \\
      \hline
      $\bz \to \etap\kz$
      & $55.4\pm \hdig 5.2\pm 4.0$
      & $68\hdd\pm 10\hdd\aer{9\hdd}{8\hdd}$ \\
      $\hsg \eta\kstarz$
      & $19.8\aer{6.5}{5.6}\pm 1.7$
      & $21.2\aer{5.4}{4.7}\pm 2.0$ \\
      $\hsg \eta\kz$ & \cc{$<\hdig 9.3$}     & \cc{$<12\hdd$} \\
      $\hsg \kz\pip\pim$
      & $47\hdd\pm 5\hdd\pm 6\hdd$
      & $50\hdd\aer{10\hdd}{\hdig 9\hdd}\pm 7\hdd$ \\
      $\hsg \kz\kp\km$
      & \cc{---}
      & $29.3\pm 3.4\pm 4.1$ \\
      $\hsg \kz\phi$
      & $\hdig 8.7\aer{1.7}{1.5}\pm 0.9$
      & $13.0\aer{6.1}{5.2}\pm 2.6$ \\
      $\hsg \ks\ks\ks$
      & \cc{---}
      & $\hdig 4.3\aer{1.6}{1.4}\pm 0.8$ \\
      $\hsg \kstarz\phi$
      & $11.1\aer{1.3}{1.2}\pm 1.1$
      & \cc{---}\\
      $\hsg \rhop\pim$
      & $28.9\pm 5.4\pm 4.3$
      & $20.8\aer{6.0}{6.3}\aer{2.8}{3.1}$ \\
      \hline
      $\bp \hsa \to \etap\kp$
      & $76.9\pm 3.5\pm 4.4$
      & $78\hdd\pm 6\hdd\pm 9$ \\
      $\hsg \eta\kp$
      & $\hdig 3.8\aer{1.8}{1.5}\pm 0.2$
      & $\hdig 5.3\aer{1.8}{1.5}\pm 0.6$ \\
      $\hsg \eta\kstarp$
      & $22.1\aer{11.1}{\hdig 9.2}\pm 3.3$
      & \cc{$<49.9$} \\
      $\hsg \omega\kp$
      & $\hdig 1.4\aer{1.3}{1.0}\pm 0.3$
      & $\hdig 9.2\aer{2.6}{2.3}\pm 1.0$ \\
      $\hsg \kstarz\pip$
      & $15.5\pm 3.4\pm 1.8$
      & \cc{---} \\
      $\hsg \kp\pip\pim$
      & $59.2\pm 4.7\pm 4.9$
      & $53.9\pm 3.1\pm 5.7$ \\
      $\hsg \kstarp\rhoz$
      & $\hdig 7.7\aer{2.1}{2.0}\pm 1.4$
      & \cc{---} \\
      $\hsg \kp\km\kp$
      & $34.7\pm 2.0\pm 1.8$
      & $33.0\pm 1.8\pm 3.2$ \\
      $\hsg \kp\phi$
      & $\hdig 9.2\pm 1.0\pm 0.8$
      & $14.6\pm 3.0\aer{2.8}{2.0}$ \\
      $\hsg \kp\ks\ks$
      & \cc{---}
      & $13.4\pm 1.9\pm 1.5$ \\
      $\hsg \kstarp\phi$
      & $12.1\aer{2.1}{1.9}\pm 1.5$
      & $11.2\aer{3.3}{2.9}\aer{1.3}{1.7}$ \\
      $\hsg \rhoz\pip$
      & $24\hdd \pm 8\hdd \pm 3\hdd$
      & $\hdig 8.0\aer{2.3}{2.0}\pm 0.7$ \\
      $\hsg \rhop\rhoz$
      & $\hdig 9.9\aer{2.6}{2.5}\pm 2.5$
      & $39\hdd \pm 11\hdd \aer{6\hdd}{5\hdd}\aer{3\hdd}{8\hdd}$ \\
      $\hsg \omega\pip$
      & $\hdig 6.6\aer{2.1}{1.8}\pm 0.7$
      & $\hdig 4.2\aer{2.0}{1.8}\pm 0.5$ \\
      $\hsg \eta\pip$
      & $\hdig 2.2\aer{1.8}{1.6}\pm 0.1$
      & $\hdig 5.2\aer{2.0}{1.7}\pm 0.6$ \\
      \hline
    \end{tabular}
  \end{center}
\end{table}
For modes with significance below three standard deviations,
90\% confidence level (C.L.) upper limits are reported.

\begin{figure}[tbp]
  \begin{center}
    \epsfig{figure=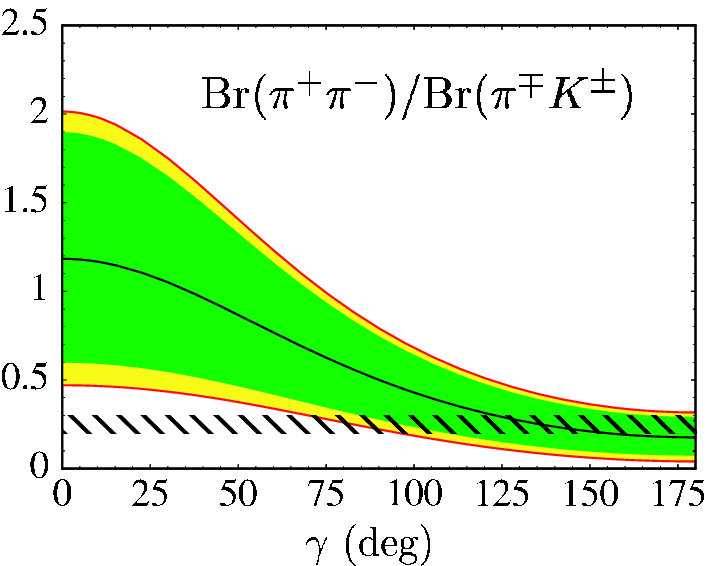,width=0.32\textwidth}
    \epsfig{figure=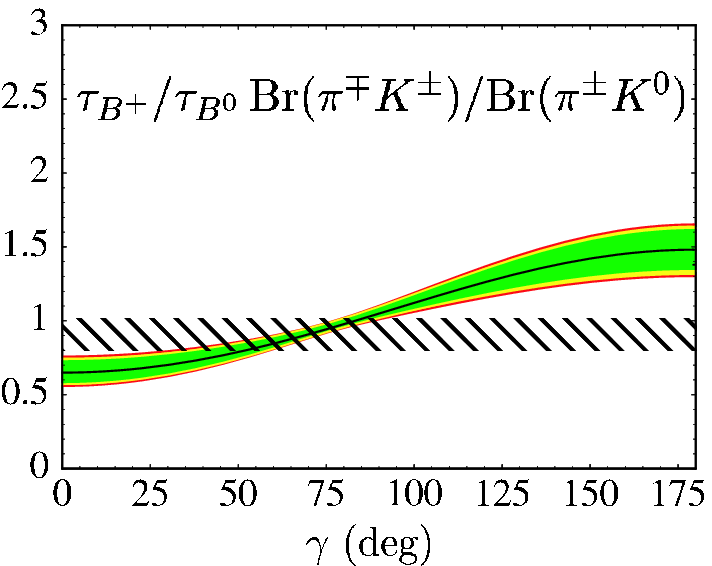,width=0.32\textwidth}
    \epsfig{figure=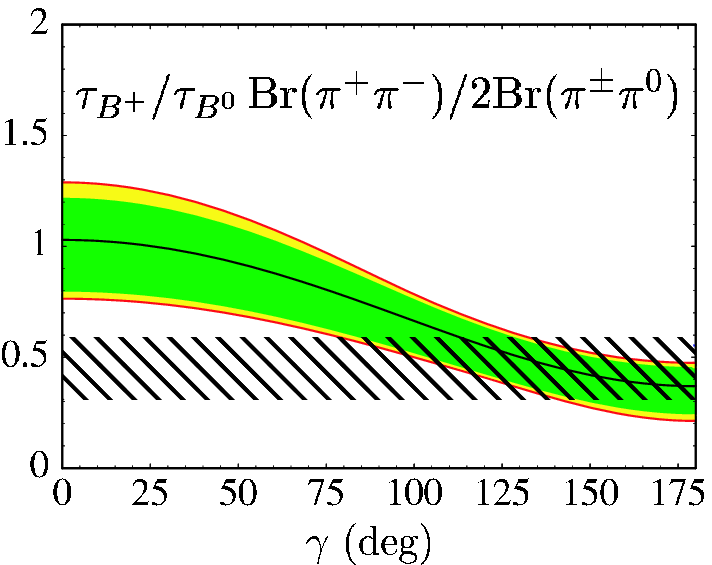,width=0.32\textwidth} \\
    \epsfig{figure=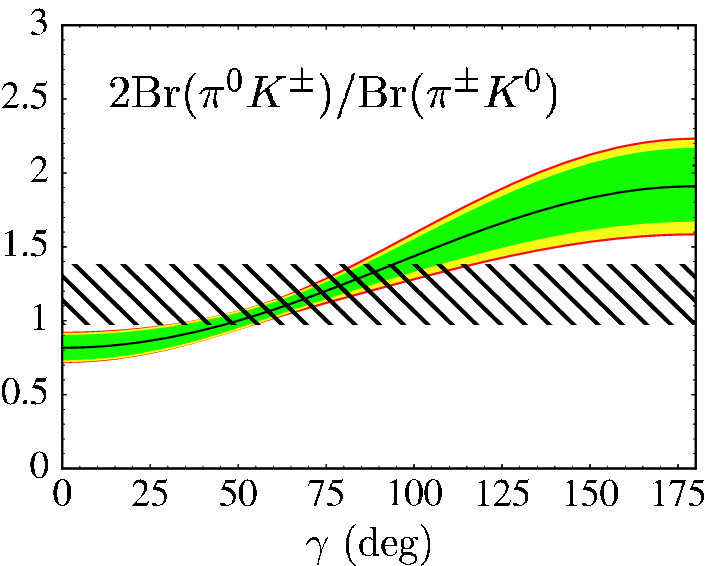,width=0.32\textwidth}
    \epsfig{figure=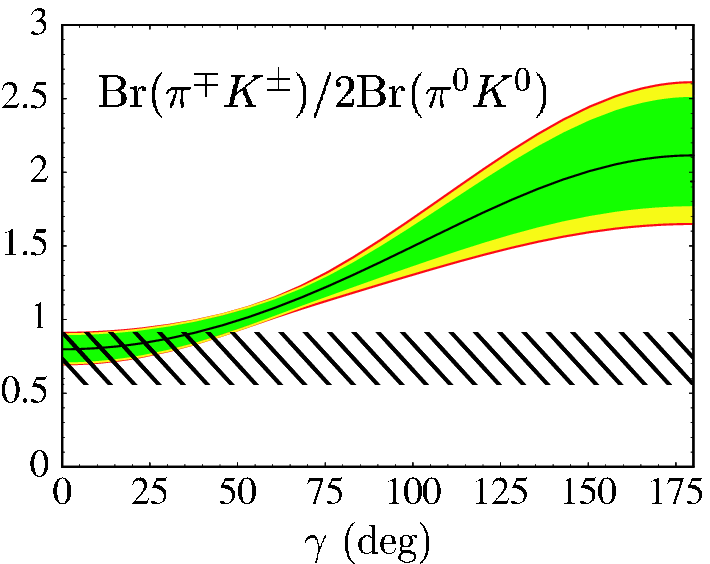,width=0.32\textwidth}
  \end{center}
  \caption{Ratios of branching fractions as functions of $\phi_3$ ($\gamma$).
    The horizontal bands show the experimental results from Belle.
    \label{fig:phi3}}
\end{figure}

The ratios of partial widths for $B \to hh$ decays
are calculated using the measurements of branching fractions from Belle
and listed in Table~\ref{tab:belle_bf_ratio}.
\begin{table}[tbp]
  \caption{Ratios of partial widths calculated
    from results of Belle. \label{tab:belle_bf_ratio}}
  \vspace{0.4cm}
  \begin{center}
    \newcommand{\m}{\hphantom{$-$}}
    \newcommand{\hdot}{\hphantom{.}}
    \newcommand{\hdig}{\hphantom{0}}
    \newcommand{\hdd}{\hdot\hdig}
    \newcommand{\cc}[1]{\multicolumn{1}{c|}{#1}}
    \newcommand{\ccc}[1]{\multicolumn{1}{|c|}{#1}}
    \newcommand{\hsa}{\hspace{+1.0mm}}
    \newcommand{\hsb}{\hspace{+0.5mm}}
    \newcommand{\hsd}{\hspace{+1.1mm}}
    \newcommand{\hse}{\hspace{+3.0mm}}
    \newcommand{\aer}[2]{\mbox{$^{\hsd+\hse #1}_{\hsd-\hse #2}$}}
    \begin{tabular}{|l|l|}
      \hline
      \ccc{Modes} & \cc{Ratio} \\
      \hline
      $\hdig \Gamma(\pip\pim) \hsa / \hdig \Gamma(\kp\pim)$
      & $0.24 \pm 0.04 \pm 0.02$ \\
      $2     \Gamma(\kp\piz)  \hsb / \hdig \Gamma(\kz\pip)$
      & $1.16 \pm 0.16 \aer{0.14}{0.11}$ \\
      $\hdig \Gamma(\kp\pim)       / \hdig \Gamma(\kz\pip)$
      & $0.91 \pm 0.09 \pm 0.06$ \\
      $\hdig \Gamma(\kp\pim)       / 2     \Gamma(\kz\piz)$
      & $0.74 \pm 0.15 \pm 0.09$ \\
      $\hdig \Gamma(\pip\pim) \hsa / 2     \Gamma(\pip\piz)$
      & $0.45 \pm 0.13 \pm 0.05$ \\
      $\hdig \Gamma(\piz\piz) \hse / \hdig \Gamma(\pip\piz)$
      & \cc{$< 0.92$} \\
      \hline
    \end{tabular}
  \end{center}
\end{table}
For the calculation of the ratio between \bz{} and \bp{} decays,
$\tau_\bp/\tau_\bz = 1.083 \pm 0.017$~\cite{PDG}
and $f_+/f_0 = 1$ are applied, where $\tau_\bp$ ($\tau_bz$)
is the lifetime of \bp{} (\bz{}) and $f_+$ ($f_0$)
is the branching fraction of $\UPS \to \bp\bm$ ($\bz\bzb$).
These ratios of branching fractions can be used
to give constraints on the weak phases.~\cite{theory_hh}
For example, the QCD factorization gives model-dependent constraint
on $\phi_3$ ($\gamma$).  Figure~\ref{fig:phi3} shows
the dependence of these ratios on $\phi_3$ ($\gamma$)
obtained from the BBNS QCD factorization approach.~\cite{bbns}
The results of ratios of branching fractions from Belle
are also displayed in Fig.~\ref{fig:phi3}.

\begin{figure}[tbp]
  \begin{center}
    \epsfig{figure=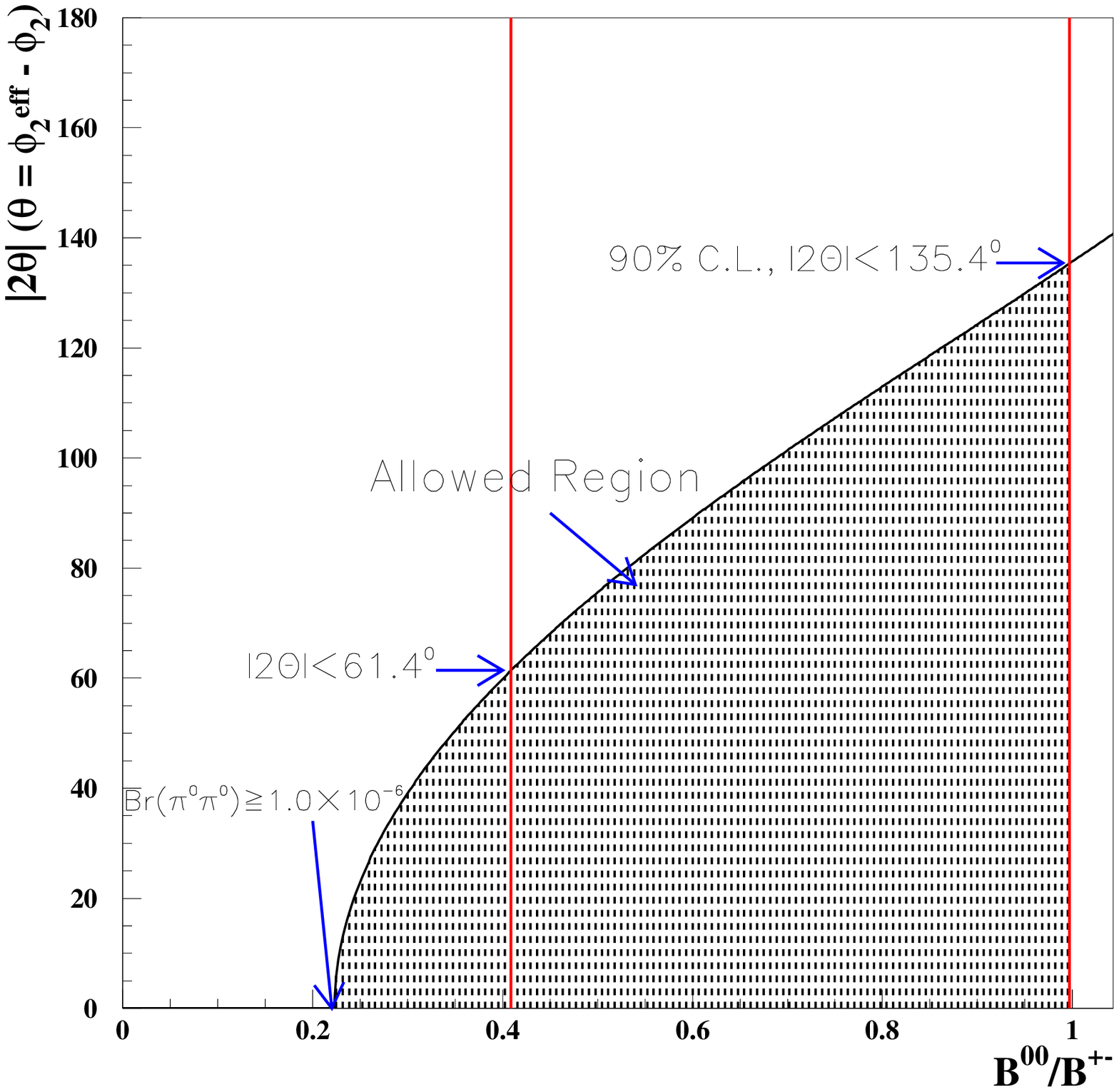,width=0.6\textwidth}
  \end{center}
  \caption{Constraint on $\phi_2$ calculated using results from Belle.
    \label{fig:phi2}}
\end{figure}

The branching fractions for $\bz \to \pip\pim$, $\bp \to \pip\piz$,
and $\bz \to \piz\piz$ can be used for the constraint
on the size of the penguin ``pollution''
in the $\phi_2$ ($\alpha$) measurement using
time-dependent asymmetry in $\bz \to \pip\pim$ decay.~\cite{theta}
The upper bound on $|2\theta| \equiv |2(\phi_2^\text{eff} - \phi_2)|$,
where $\phi_2^\text{eff}$ is the measured parameter
from the $CP$ asymmetry in the time-evolution of $\bz \to \pip\pim$ decay,
is calculated using the results from Belle:
$R \equiv \Br(\piz\piz)/\Br(\pip\pim) = 0.41 < 1.00 \text{ (90\% C.L.)}$
and $A_{\pi\pi} = 0.57$ obtained with the constraint
to the physically allowed region.~\cite{belle_pipi}
The obtained allowed region for $|2\theta|$ and $R$
is shown in Fig.~\ref{fig:phi2},
and the typical values of the upper limit on $|2\theta|$ are
\begin{alignat}{2}
  |2\theta| &< 61.4\degree & \text{ (for $R = 0.41$)} \\
  &< 135.4\degree & \text{ (for $R = 1.00$)} .
\end{alignat}
The lower bound on $\Br(\piz\piz)$ can be estimated
to be $\Br(\piz\piz) \ge 1.0 \times 10^{-6}$
from Fig.~\ref{fig:phi2}.

\subsection{Direct $CP$ Violation}

\begin{figure}[tbp]
  \begin{center}
    \epsfig{figure=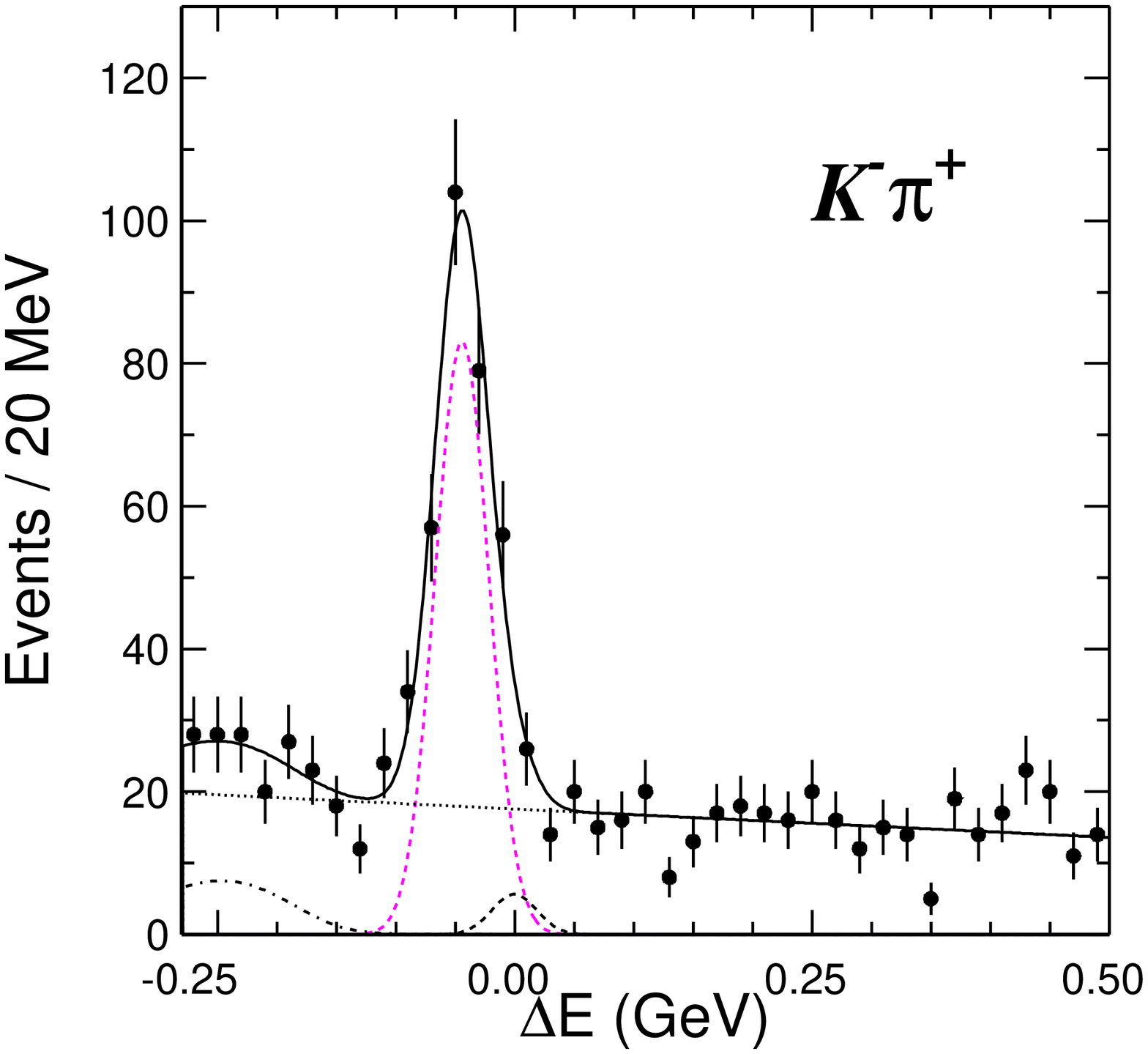,width=0.3\textwidth}
    \epsfig{figure=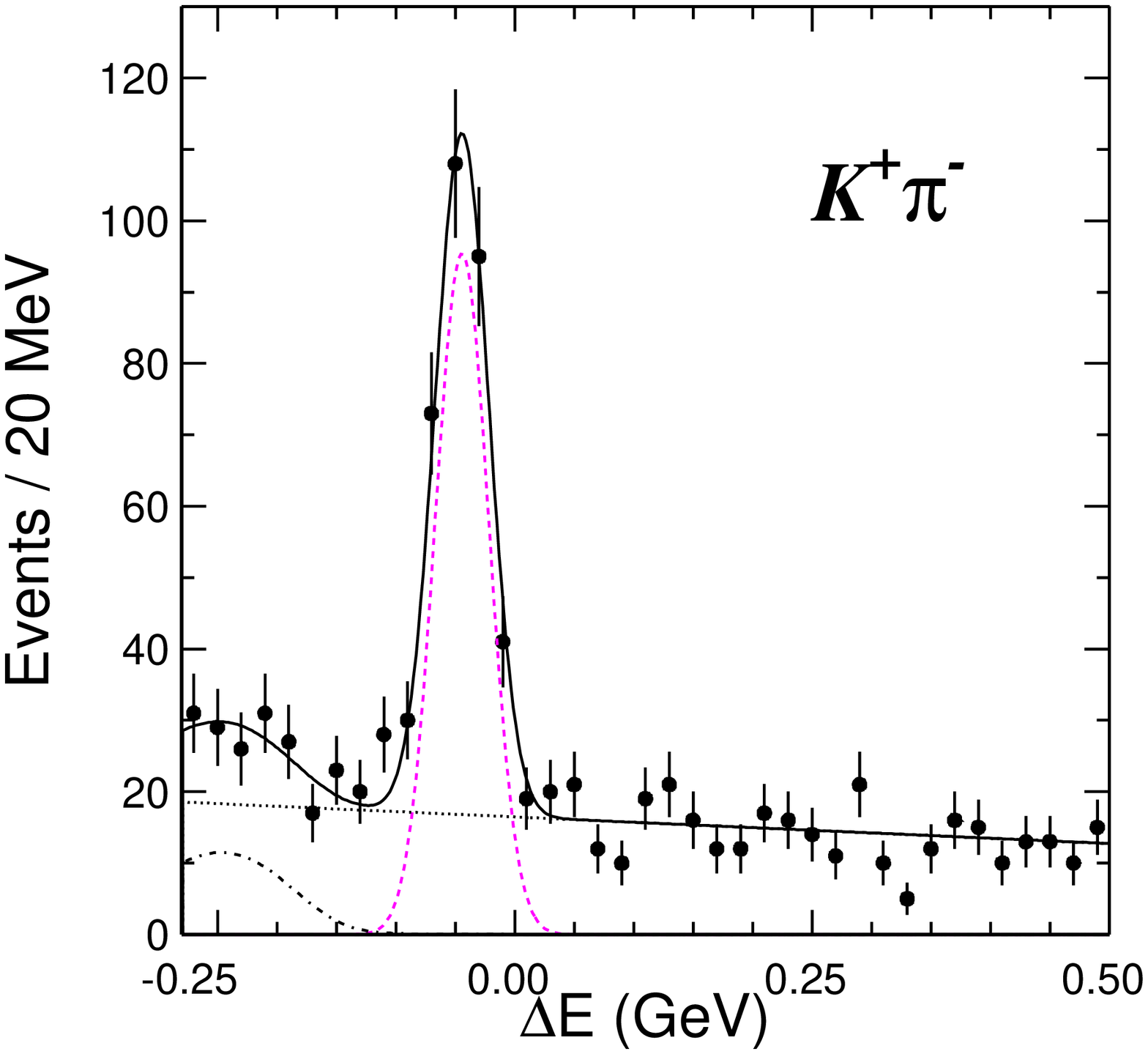,width=0.3\textwidth} \\
    \vspace*{-1ex}
    \makebox[0.6\textwidth][l]{(a) $\bz \to \kp\pim$} \\
    \vspace*{4ex}
    \epsfig{figure=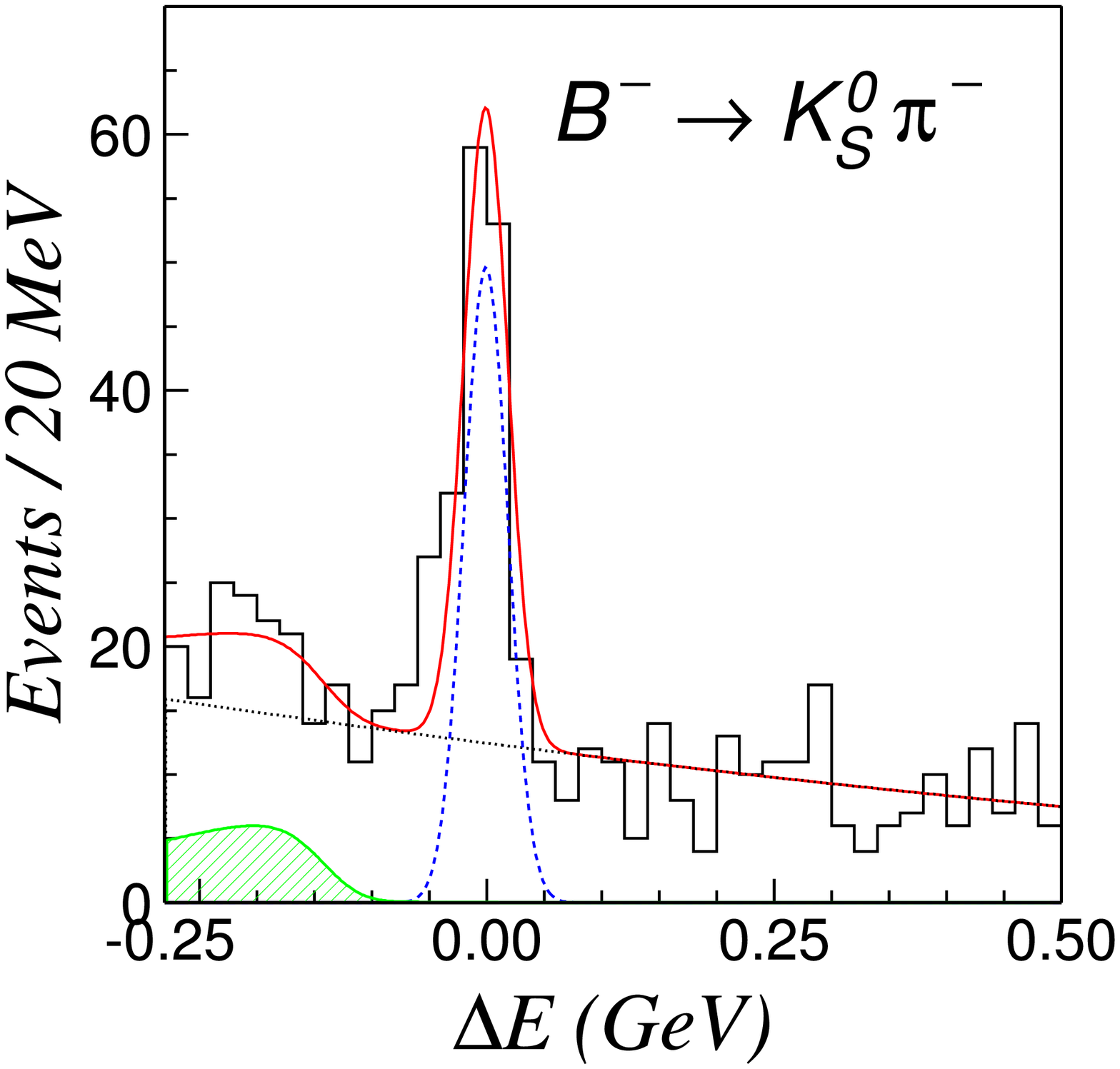,width=0.3\textwidth}
    \epsfig{figure=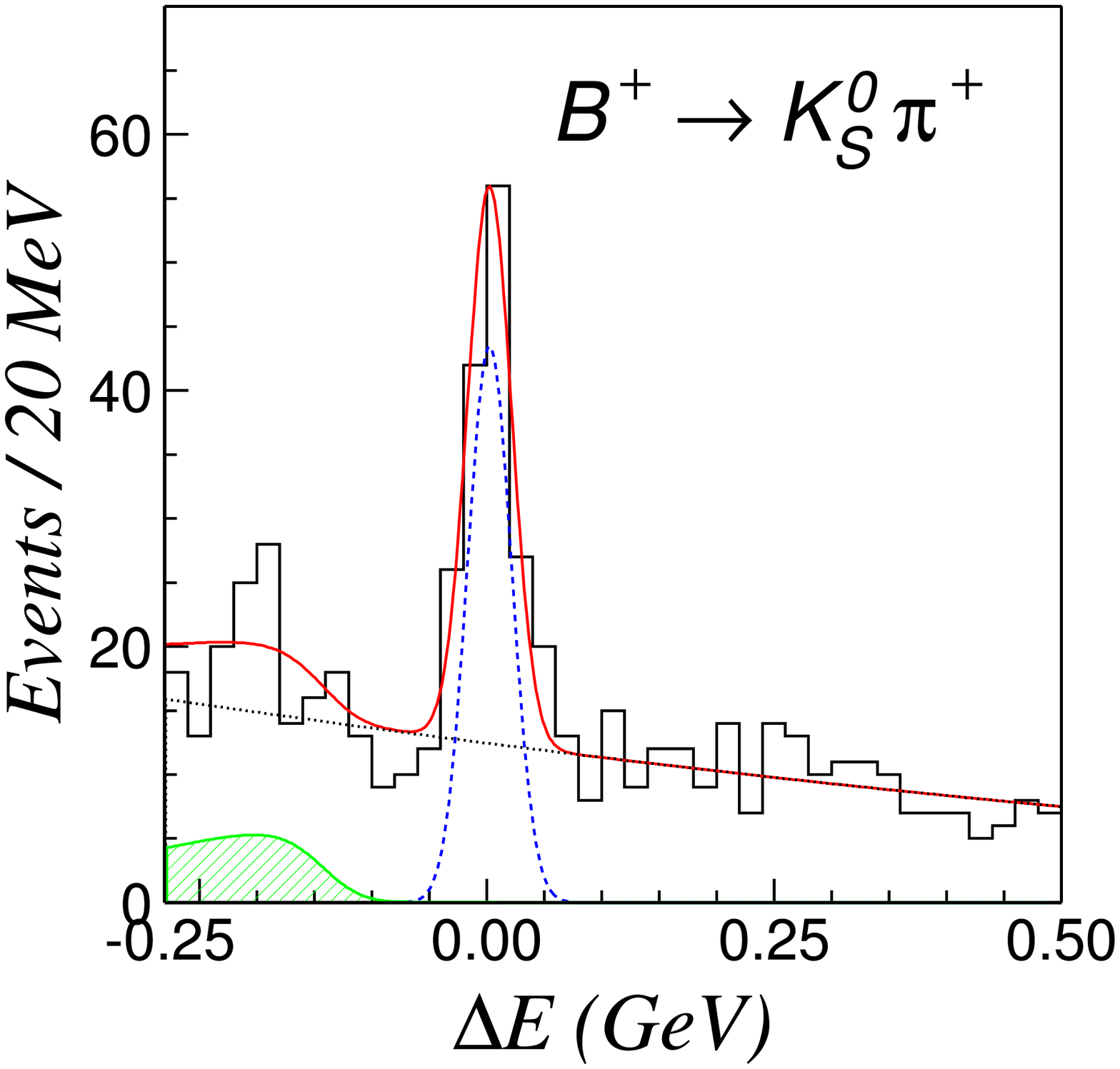,width=0.3\textwidth} \\
    \makebox[0.6\textwidth][l]{(b) $\bp \to \ks\pip$} \\
    \vspace*{1ex}
    \epsfig{figure=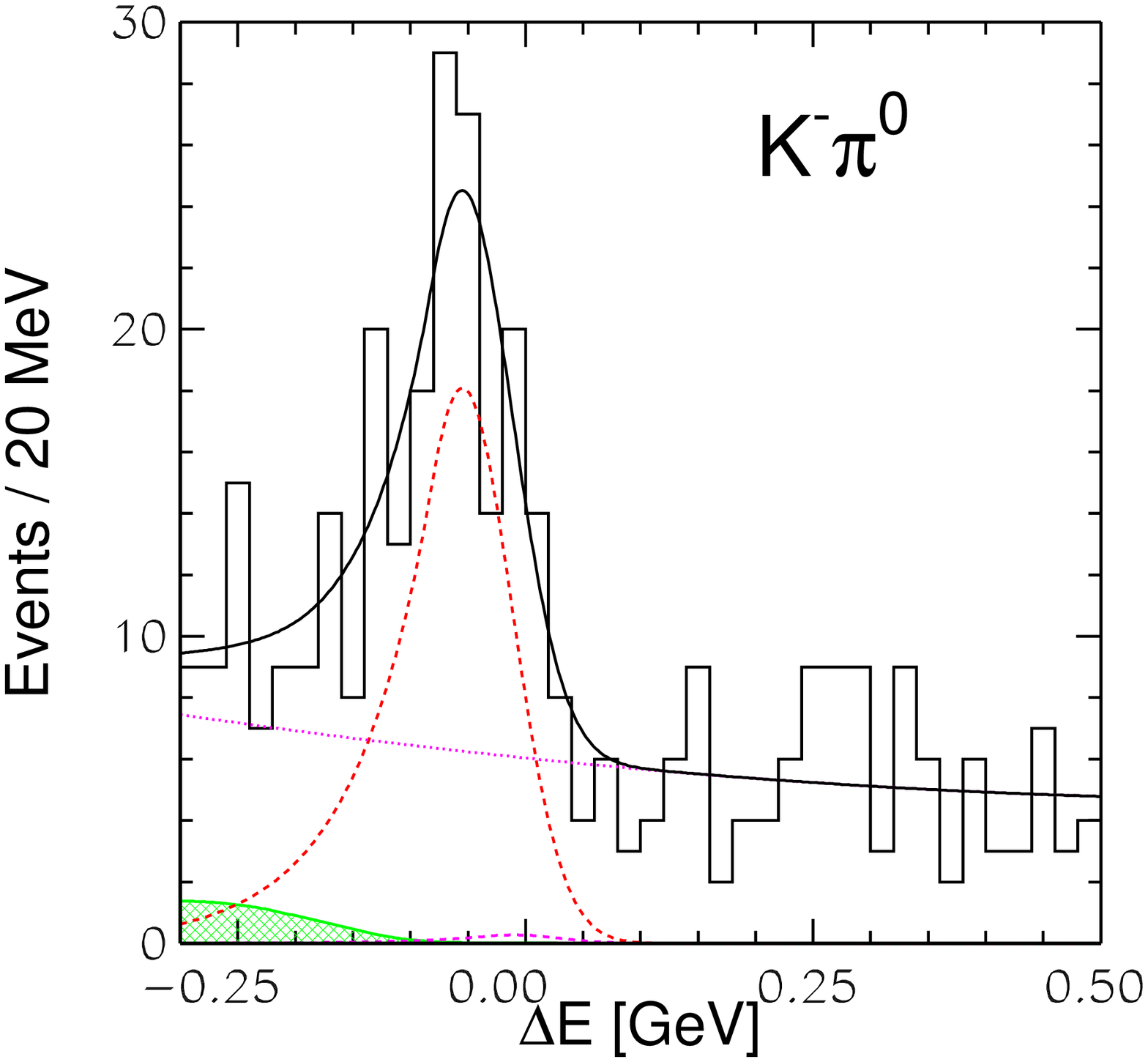,width=0.3\textwidth}
    \epsfig{figure=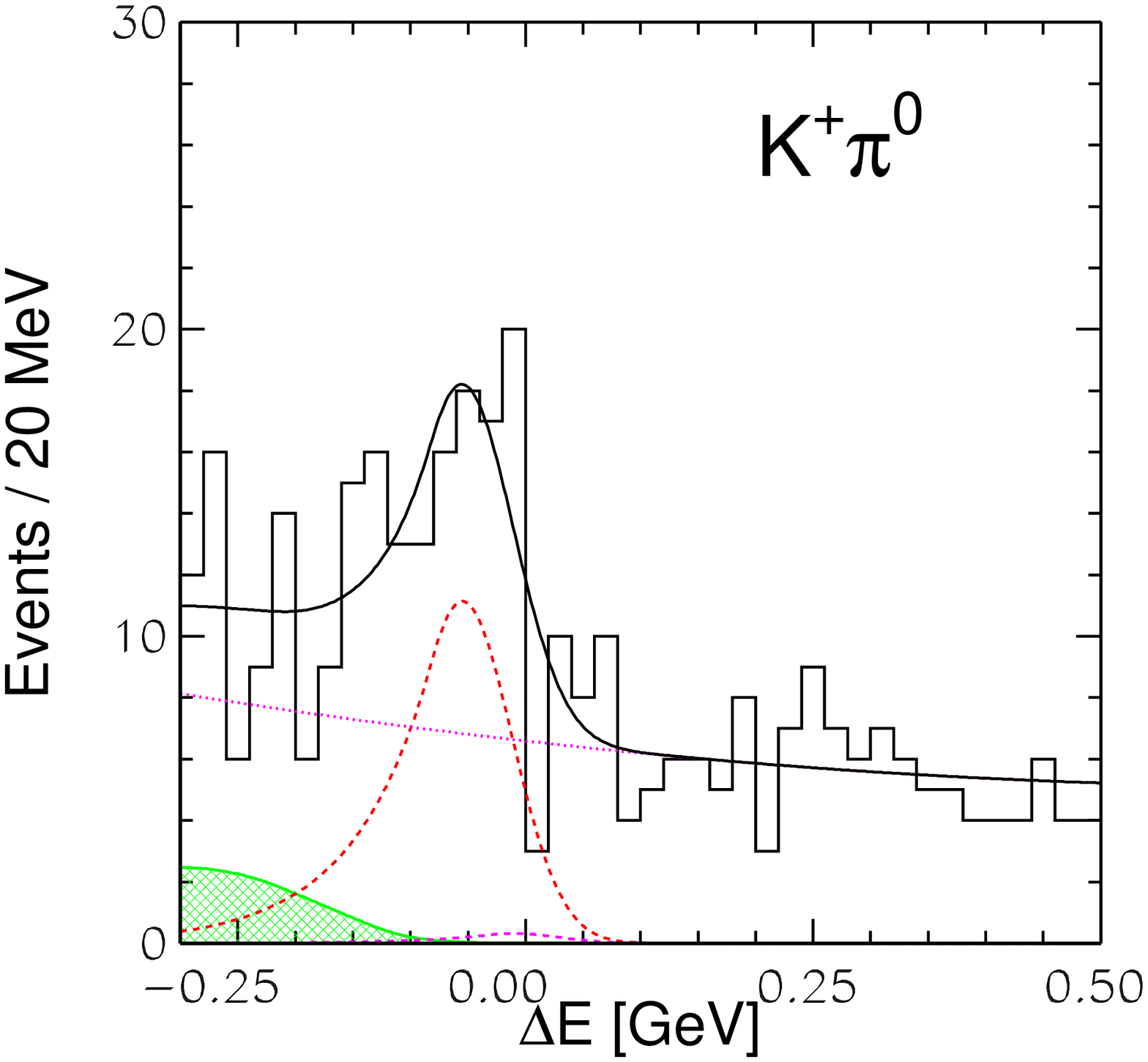,width=0.3\textwidth} \\
    \vspace*{-1ex}
    \makebox[0.6\textwidth][l]{(c) $\bp \to \kp\piz$} \\
    \vspace*{1ex}
    \epsfig{figure=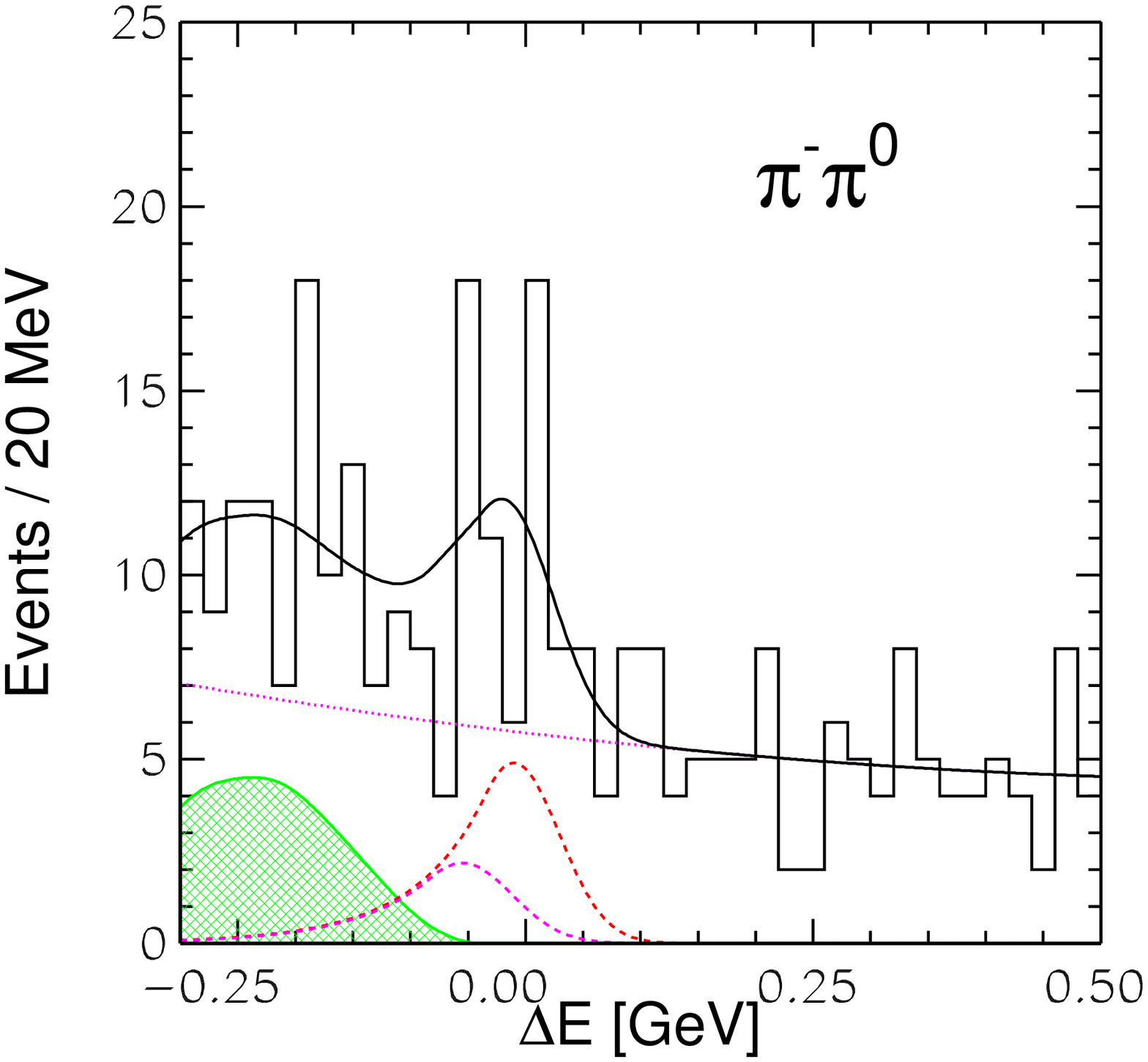,width=0.3\textwidth}
    \epsfig{figure=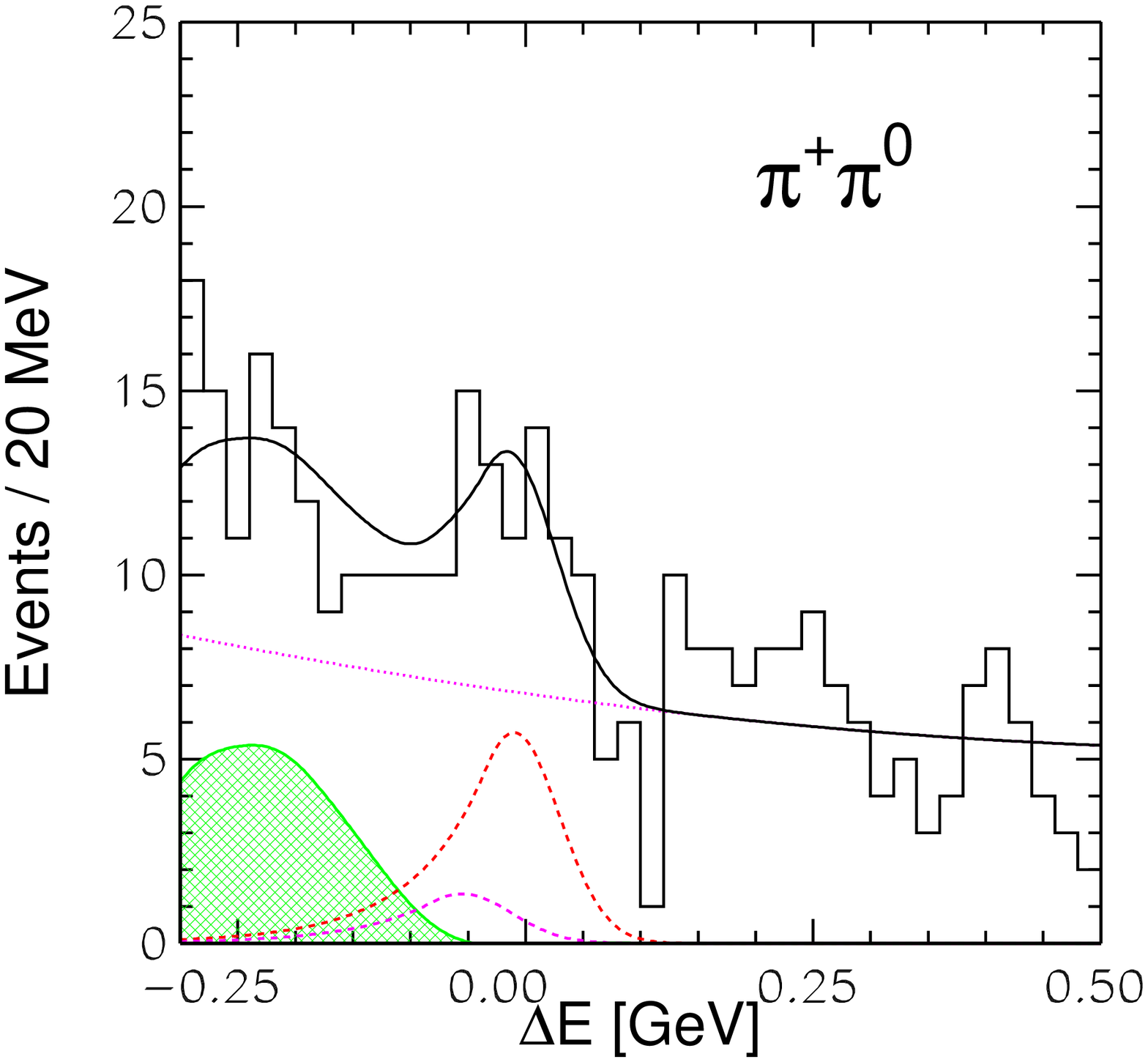,width=0.3\textwidth} \\
    \vspace*{-1ex}
    \makebox[0.6\textwidth][l]{(d) $\bp \to \pip\piz$}
  \end{center}
  \caption{Distributions of \dE{} and fit results from Belle
    for (a) $\bz \to \kp\pim$, (b) $\bp \to \ks\pip$, (c) $\bp \to \kp\piz$,
    and (d) $\bp \to \pip\piz$ divided into $\bzb/\bm$ (left) and
    $\bz/\bp$ (right). \label{fig:belle_acp}}
\end{figure}

Belle measures the partial-rate asymmetry \Acp{}
by fitting the \dE{} distributions and extracting signal yields
separately for \bz{} (\bp{}) and \bzb{} (\bm{}).
Figure~\ref{fig:belle_acp} shows the \dE{} distributions
separately for \bz{} (\bp{}) and \bzb{} (\bm{}) modes
for $B \to hh$ decays from Belle.
The fitting results and partial-rate asymmetries from Belle are
listed in Table~\ref{tab:acp_belle}.
\begin{table}[tbp]
  \caption{Number of signal events for $\bbar$ (\bzb{} or \bm{})
    and $B$ (\bz{} or \bp{}), and partial-rate asymmetry \Acp{}
    for $B \to hh$ modes from Belle.
    90\% confidence interval (C.I.) for \Acp{} is also shown.
    \label{tab:acp_belle}}
  \vspace{0.4cm}
  \begin{center}
    \newcommand{\m}{\hphantom{$-$}}
    \newcommand{\hdot}{\hphantom{.}}
    \newcommand{\hdig}{\hphantom{0}}
    \newcommand{\hdd}{\hdot\hdig}
    \newcommand{\cc}[1]{\multicolumn{1}{c|}{#1}}
    \newcommand{\ccc}[1]{\multicolumn{1}{|c|}{#1}}
    \newcommand{\hsa}{\hspace{-0.7mm}}
    \newcommand{\hsb}{\hspace{+1.6mm}}
    \newcommand{\hsd}{\hspace{+1.5mm}}
    \newcommand{\hse}{\hspace{+3.0mm}}
    \newcommand{\aer}[2]{\mbox{$^{\hsd+\hse #1}_{\hsd-\hse #2}$}}
    \begin{tabular}{|l|l|l|ll|}
      \hline
      \ccc{Mode} & \cc{$\Ns(\bbar)$} & \cc{$\Ns(B)$}
      & \multicolumn{1}{c}{$\Acp$} & \cc{($90\%$ C.I.)} \\
      \hline
      $\kp\pim$
      & $235.4\aer{19.8}{19.1}$
      & $270.2\aer{19.7}{18.9}$
      & $-0.07\pm 0.06\pm 0.01$
      & ($-0.18 < \Acp < 0.04$) \\
      $\kp\piz$
      & $122.0\pm 15.8$
      & $\hdig 76.5\pm 14.5$
      & \m$0.23\pm 0.11\aer{0.01}{0.04}$
      & ($-0.01 < \Acp < 0.42$) \\
      $\kz\pip$
      & $119.1\aer{13.8}{13.1}$
      & $104.4\aer{13.2}{12.5}$
      & \m$ 0.07\aer{0.09}{0.08}\aer{0.01}{0.03}$
      & ($-0.10 < \Acp < 0.22$) \\
      $\pip\piz$
      & $\hdig 31.2\pm 11.9$
      & $\hdig 41.3\pm 12.7$
      & $-0.14\pm 0.24\aer{0.05}{0.04}$
      & ($-0.57 < \Acp < 0.30$) \\
      \hline
    \end{tabular}
  \end{center}
\end{table}

\babar{} uses an unbinned maximum likelihood fit
to determine the partial-rate asymmetry \Acp{}.
The input parameters are the same as those
used in the measurement of branching fractions.
Table~\ref{tab:acp_babar} lists \Acp{} for $B \to hh$ modes
from \babar{}.~\cite{babar_prl,babar_ichep02}
\begin{table}[tbp]
  \caption{Summary of \Acp{} for $B \to hh$ modes from \babar{}.
    \label{tab:acp_babar}}
  \vspace{0.4cm}
  \begin{center}
    \newcommand{\m}{\hphantom{$-$}}
    \newcommand{\hdot}{\hphantom{.}}
    \newcommand{\hdig}{\hphantom{0}}
    \newcommand{\hdd}{\hdot\hdig}
    \newcommand{\cc}[1]{\multicolumn{1}{c|}{#1}}
    \newcommand{\ccc}[1]{\multicolumn{1}{|c|}{#1}}
    \newcommand{\hsa}{\hspace{-0.7mm}}
    \newcommand{\hsb}{\hspace{+1.6mm}}
    \newcommand{\hsd}{\hspace{+1.1mm}}
    \newcommand{\hse}{\hspace{+2.0mm}}
    \newcommand{\aer}[2]{\mbox{$^{\hsd+\hse #1}_{\hsd-\hse #2}$}}
    \begin{tabular}{|l|l|}
      \hline
      \ccc{Mode} & \cc{$\Acp$} \\
      \hline
      \kp\pim
      & $-0.102 \pm 0.050 \pm 0.016$ \\
      \kp\piz
      & $-0.09 \hdig \pm 0.09 \hdig \pm 0.01$ \\
      \kz\pip
      & $-0.17 \hdig \pm 0.10 \hdig \pm 0.02$ \\
      \kz\piz
      & \m $ 0.03 \hdig \pm 0.36 \hdig \pm 0.09$ \\
      \pip\piz
      & $-0.03 \hdig \aer{0.18 \hdig}{0.17 \hdig} \hspace*{1pt} \hdig \pm 0.02$ \\
      \hline
    \end{tabular}
  \end{center}
\end{table}

\begin{figure}[tbp]
  \begin{center}
    \epsfig{figure=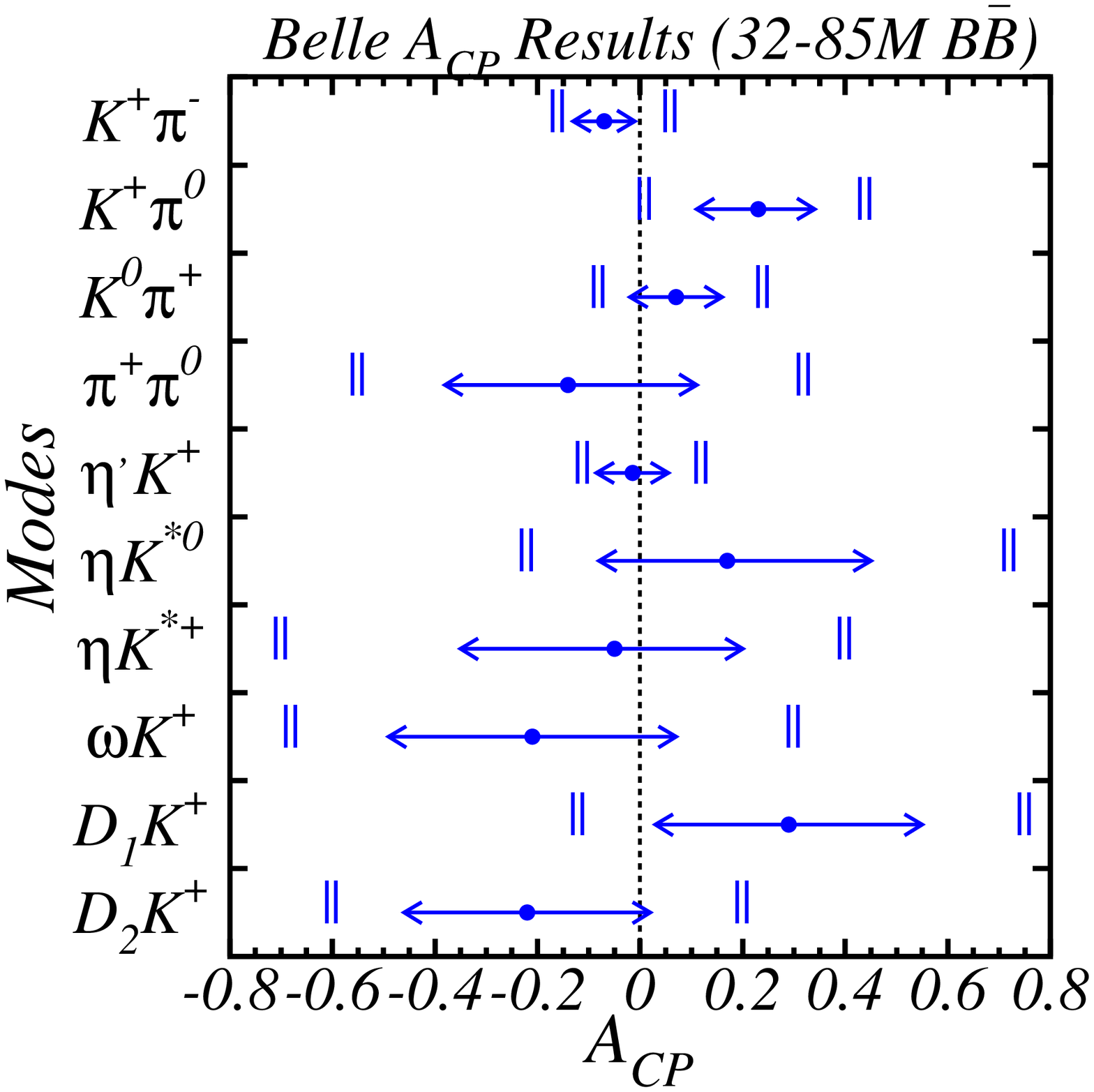,height=0.5\textwidth}
    \epsfig{figure=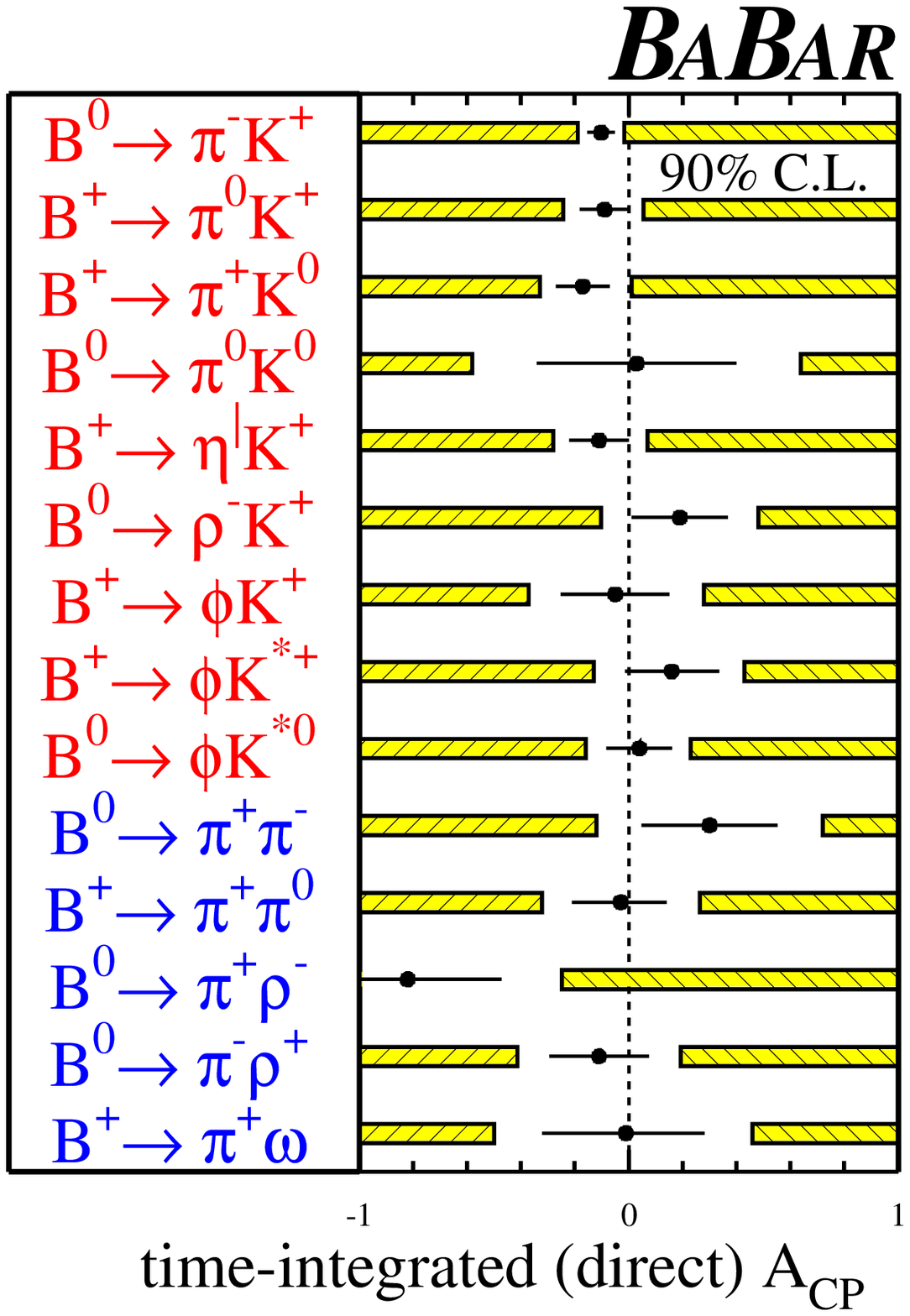,height=0.5\textwidth}
  \end{center}
  \caption{Summary of \Acp{} results from Belle (left) and \babar{} (right).
    The point, arrows (line), and bar represents the \Acp{},
    errors, and 90\% confidence intervals, respectively. \label{fig:acp_sum}}
\end{figure}

Figure~\ref{fig:acp_sum} shows the summary plots of \Acp{}
for rare hadronic $B$ decays from Belle and \babar{}
including the modes other than $B \to hh$.~\cite{rare_hfag}

\section{Summary}

The $B$-factory experiments provide the most precise measurements
of branching fractions and partial-rate asymmetries
for rare hadronic $B$ decays.
The measured partial-rate asymmetries are consistent with zero
with the current statistics.
The statistical precisions for these measurements
have reached below 10\% level in several decay modes.

\end{document}